\documentclass[pdflatex]{sn-jnl}

\usepackage{graphicx}%
\usepackage{multirow}%
\usepackage{amsmath,amssymb,amsfonts}%
\usepackage{amsthm}%
\usepackage{mathrsfs}%
\usepackage{tabularx}
\usepackage[title]{appendix}%
\usepackage{xcolor}%
\usepackage{textcomp}%
\usepackage{manyfoot}%
\usepackage{booktabs}%
\usepackage{algorithm}%
\usepackage{algorithmicx}%
\usepackage{algpseudocode}%
\usepackage{listings}%
\usepackage[utf8]{inputenc} 
\usepackage[T1]{fontenc}    
\usepackage{hyperref}       
\usepackage{url}            
\usepackage{amsfonts}       
\usepackage{nicefrac}       
\usepackage{microtype}      
\usepackage{lipsum}		
\usepackage{doi}
\usepackage{verbatim}
\usepackage{caption}
\usepackage{array}       
\usepackage{colortbl}  
\usepackage[normalem]{ulem}

\newcommand{\exchange}[2]{#2}

\usepackage{wrapfig}
\usepackage{diagbox} 
\usepackage{float} 
\theoremstyle{thmstyleone}%
\theoremstyle{thmstyletwo}%

\theoremstyle{thmstylethree}%

\newcommand{\DB}{\textit{DishBrain} }
\newcommand{\iv}{\textit{in vitro} }
\newcommand{\TV}{TAVRNN}

\usepackage{titlesec}

\titleformat{\paragraph}
  {\normalfont\normalsize\bfseries} 
  {} 
  {0pt} 
  {} 

\titlespacing*{\paragraph}
  {0pt}  
  {6pt plus 2pt minus 2pt}  
  {4pt}  

\begin{document}

\title{Graph-Based Representation Learning of Neuronal Dynamics and Behavior}




\author[1]{\fnm{Moein} \sur{Khajehnejad}}
\equalcont{These authors contributed equally to this work.}

\author[2]{\fnm{Forough} \sur{Habibollahi}}
\equalcont{These authors contributed equally to this work.}

\author[3]{\fnm{Ahmad} \sur{Khajehnejad}}

\author[4]{\fnm{Chris} \sur{French}}

\author[2]{\fnm{Brett J.} \sur{Kagan}}

\author*[1,5,6]{\fnm{Adeel} \sur{Razi}}\email{adeel.razi@monash.edu}

\affil[1]{\orgdiv{Turner Institute for Brain and Mental Health}, \orgname{School of Psychological Sciences, Monash University}, \orgaddress{\city{Melbourne}, \country{Australia}}}

\affil[2]{\orgdiv{Cortical Labs}, \orgaddress{\city{Melbourne}, \country{Australia}}}

\affil[3]{\orgdiv{Department of Computer Science}, \orgname{The University of British Columbia}, \orgaddress{\city{Vancouver}, \country{Canada}}}

\affil[4]{\orgdiv{Neural Dynamics Laboratory, Department of Medicine}, \orgname{The University of Melbourne}, \orgaddress{\city{Melbourne}, \country{Australia}}}

\affil[5]{\orgdiv{Wellcome Centre for Human Neuroimaging}, \orgname{University College London}, \orgaddress{\city{London}, \country{United Kingdom}}}
\affil[6]{\orgname{CIFAR Azrieli Global Scholars Program}, \orgaddress{\city{Toronto}, \country{Canada}}}

\abstract{
Understanding how neuronal networks reorganize in response to external stimuli and give rise to behavior is a central challenge in neuroscience and artificial intelligence. However, existing methods often fail to capture the evolving structure of neural connectivity in ways that capture its relationship to behavior, especially in dynamic, uncertain, or high-dimensional settings with sufficient resolution or interpretability. We introduce the Temporal Attention-enhanced Variational Graph Recurrent Neural Network (TAVRNN), a novel framework that models time-varying neuronal connectivity by integrating probabilistic graph learning with temporal attention mechanisms. TAVRNN learns latent dynamics at the single-unit level while maintaining interpretable population-level representations, to identify key connectivity patterns linked to behavior. TAVRNN generalizes across diverse neural systems and modalities, demonstrating state-of-the-art classification and clustering performance. We validate TAVRNN on three diverse datasets: (1) electrophysiological data from a freely behaving rat, (2) primate somatosensory cortex recordings during a reaching task, and (3) biological neurons in the \DB platform interacting with a virtual game environment. Our method outperforms state-of-the-art dynamic embedding techniques, revealing previously unreported relationships between adaptive behavior and the evolving topological organization of neural networks. These findings demonstrate that TAVRNN offers a powerful and generalizable approach for modeling neural dynamics across experimental and synthetic biological systems. Its architecture is modality-agnostic and scalable, making it applicable across a wide range of neural recording platforms and behavioral paradigms.}

\maketitle
\section{Main}
Artificial intelligence has long drawn inspiration from biological systems, with neuroscience providing foundational insights into architectures and learning rules, from the McCulloch–Pitts neuron and connectionist models to contemporary proposals for NeuroAI \cite{richards2019deep,zador2023catalyzing}. As advances in neural recording now allow simultaneous measurement of large-scale neuronal activity, a critical challenge emerges: how to link high-dimensional neural dynamics to computational processes and behavior in a way that is both scalable and interpretable, a goal that has received significant recent attention \cite{sani2024dissociative,vahidi2024modeling,azabou2023unified}. 


Conventional graph-based approaches have facilitated the study of brain connectivity by embedding nodes (neurons or regions) into a lower-dimensional space \cite{donnat2018learning}. These representations enable downstream analyses such as classification, clustering, or behavior prediction.
However, most current approaches assume static networks, applying fixed embeddings to network snapshots while disregarding the dynamic nature of neuronal connectivity \cite{armandpour2019robust,grover2016node2vec,perozzi2014deepwalk,ribeiro2017struc2vec,tang2015line,khajehnejad2019simnet}. This oversimplification fails to capture the evolving properties of neuronal networks, particularly during learning, where connectivity and function evolve continuously \cite{lurie2020questions}.
Although recent work has extended network embedding techniques to network graphs \cite{goyal2018dyngem,zhou2018dynamic,trivedi2019dyrep,li2017attributed}, these approaches typically represent each node with a deterministic low-dimensional vector \cite{bojchevski2017deep}, omitting uncertainty and variability intrinsic to neural systems. Probabilistic methods, such as Variational Graph Recurrent Neural Networks (VGRNNs) \cite{hajiramezanali2019variational, yap2023deep}, introduced stochasticity into node embeddings, but often lack mechanisms to handle the heterogeneous and context-dependent nature of temporal dependencies.

Here, we present the Temporal Attention-enhanced Variational Graph Recurrent Neural Network (\TV), a novel framework that addresses these limitations by combining three key innovations: (1) variational inference to model uncertainty in node representations, (2) temporal attention to learn variable-lag dependencies across time, and (3) graph-based recurrent structure to integrate topological and dynamical information. Together, these components enable \TV{} to model how neural activity evolves over time in conjunction with learning and behavior.
By incorporating these methodological advancements, \TV{} enables a unified analysis of temporal neural dynamics and evolving network structure. Unlike prior studies that isolate neuronal signal dynamics from network structure \cite{schneider2023learnable,manley2024simultaneous,khajehnejad2020neural}, our approach integrates both individual neuron activity and evolving functional connectivity. This design allows the model to learn context-sensitive embeddings of individual neurons, while also capturing how these neurons interact and reorganize at the network level during learning.

We apply \TV{} across three neural datasets that differ in biological substrate, measurement modality, and behavioral complexity: (i) multicellular spiking activity recorded using bilaterally implanted silicon probes in the CA1 region of the hippocampus in freely navigating rats \cite{grosmark2016diversity}, (ii) electrophysiological recordings from primate somatosensory cortex during goal-directed reaching \cite{chowdhury2020area}, and (iii) the DishBrain system \cite{kagan2022vitro}, where living neurons interact with a virtual game `Pong' via closed-loop stimulation.
These experiments span a range of cognitive and sensorimotor tasks, offering a robust test of the model’s generality.
Furthermore, our model's architecture accommodates variable-length time windows, enabling adaptability to dynamic temporal structures commonly observed in neurophysiological recordings. It operates effectively in both supervised and unsupervised settings, making it suitable for both predictive modeling and exploratory data analysis.

Importantly, \TV{} seamlessly integrates multi-subject and multi-trial datasets, and is modality-agnostic—compatible with a broad range of neural recording technologies, including silicon probes, extracellular electrophysiology, Neuropixels, and fMRI. This versatility enables multimodal analysis across diverse behavioral paradigms and species, and supports investigation of neural activity at both cellular and population levels.  These features collectively position \TV{} as a powerful tool for discovering interpretable, behaviorally relevant representations in dynamic, high-dimensional, and complex neural systems.

\section{Results}
\label{results}

We evaluate \TV{} using three well-characterized \textit{in vivo} datasets: hippocampal multicellular electrophysiological data from a freely behaving rat, primate somatosensory cortex electrophysiological recordings, and in vitro neurons playing a game in virtual world embodied using a close-loop feedback system ( Figs. \ref{Schematic_rat}a, \ref{Schematic_monkey}a, and \ref{Schematic_DB}a). In the first two datasets, where ground truth labels correspond to the rat’s position on the track or the target/hand movement direction in the reaching task for the primate, \TV{} effectively distinguished neural activity patterns associated with different behavioral states and was capable of classifying them with high accuracy levels (Figs. \ref{Schematic_rat}d, and \ref{Schematic_monkey}d). It achieved high classification accuracy, aligning with—and in some cases surpassing—the performance of established methods such as both CEBRA-Time and CEBRA-Behavior variants \cite{schneider2023learnable} (see Figs. \ref{Schematic_rat}c, \ref{Schematic_monkey}c, and Supplementary Materials \ref{CEBRA_setting} for more details on the implementation settings).

To further assess the model’s ability to extract meaningful neural representations in a richer, less characterized dataset, we applied TAVRNN, to a third dataset, to \iv electrophysiological recordings from the \DB system, where biological neurons interact with a simulated game environment. Unlike the \textit{in vivo} datasets, where labeled behavioral states provide a direct learning signal, the \DB dataset lacks explicit ground truth labels, requiring the model to capture latent links between neural activity and emergent behavioral patterns in an unsupervised manner. Notably, TAVRNN uncovered a previously unreported relationship between high game performance and the alignment of sensory and motor subregion activity, providing novel insights into the dynamic reorganization of neuronal networks during adaptive learning.

\subsection{Rat hippocampus dataset}
\begin{figure}
  \centering
  \includegraphics[width = 1\textwidth]{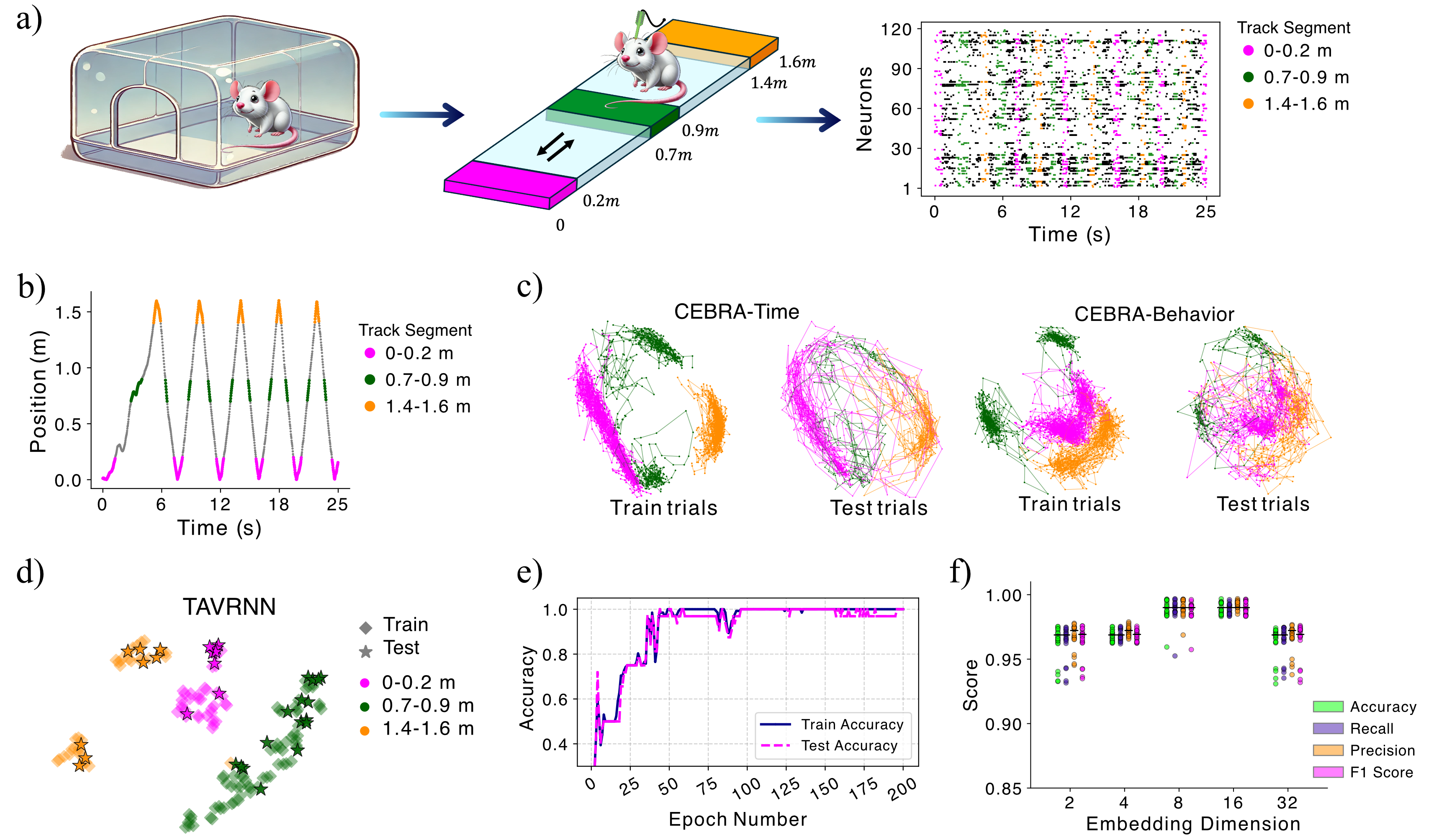}  
  \caption{\small{\textbf{a)} Schematic of rat hippocampus data collection and sample neuronal activity recorded over a 25-second window during traversal of a 1.6 m track. The track’s middle and both ends, along with the corresponding neuronal spikes recorded during these segments, are color-coded to match \textbf{a)} the behavioral plot, which illustrates the rat’s position. 
  \textbf{c)} Embedded trajectories of all 132 train and 32 test trials at the population level during time windows when the rat traversed each of the colored track segments, using CEBRA-Time and CEBRA-Behavior. While both methods exhibit trajectory separation for the three segments in train trials, the distinction between time windows with different behavioral labels disappears in unseen test trials when examining population-level embeddings (trajectories over time).
  \textbf{d)} Low-dimensional representation of neuronal activity during time windows when the rat traversed the three track segments, using \TV{}. Train trials (diamonds) and test trials (stars) largely overlap. In each time window, all single neurons are embedded into a lower-dimensional space using TAVRNN. We visualize the evolution of a summary node that captures the temporal context across all neurons, plotted for all 132 train trials and 32 test trials. \textbf{e)} Train and test set accuracy across training epochs using the \TV{} method. The train accuracy (solid blue line) and test accuracy (dashed magenta line) demonstrate convergence, with high performance achieved for both sets after sufficient training epochs. \textbf{f)} Performance metrics (accuracy, recall, precision, and F1 score) across different embedding dimensions for the \TV{} method. Each metric is represented by a distinct color, with individual points indicating the scores from 20 independent runs of the model. The embedding dimension was set to 8 for both \TV{} and CEBRA variants, with 2D visualizations used in this figure for all results.}}
  \label{Schematic_rat}
  
\end{figure}

We used the dataset from \cite{grosmark2016diversity}, consisting of multicellular recordings from 120 putative pyramidal neurons in the CA1 hippocampal subfield of a male Long–Evans rat using silicon probes. The rat ran on a 1.6-meter linear track, receiving water rewards at both ends (Fig. \ref{Schematic_rat}a). The rat’s position on the track was simultaneously recorded (Fig. \ref{Schematic_rat}b) and this data served as ground truth to validate \TV{} in a downstream classification task (see Fig. \ref{Schematic_rat}d). This task aimed to link population neuronal activity to the rat’s position on the track, a feature thought to be encoded by place cells in the hippocampus based on previous evidence \cite{o1971hippocampus}.

For this dataset, we used binary spiking data from 120 neurons across 10,178 time points at 40 Hz. We selected time windows of spiking activity when the rat was within the first, middle, and last 0.2 meters of the track, yielding 164 crossings (Fig. \ref{Schematic_rat}a,b). These varying length time windows were subsequently labeled into three classes for the downstream classification task. Of the 164 crossing windows, 132 were used for training, while the remaining 32 were reserved for testing. To ensure that the covariance matrix is not ill-conditioned in these time windows, leveraging the Marchenko-Pastur distribution \cite{marchenko1967distribution,bickel2008covariance,fan2010selective}, we compared it to a shuffled control, preserving neuron identity while shuffling time points independently. This process was repeated 1000 times to estimate confidence intervals, considering only correlations beyond the 95\% confidence bounds in the analysis. For further details, see Supplementary Materials \ref{marchenko}.
As the final step, we constructed a network adjacency matrix for each window, representing functional connectivity with zero-lag Pearson correlations as edges and 120 neurons as nodes. The choice of Pearson correlation for connectivity inference was guided by graph kernel selection (see Supplementary Materials \ref{conn_inference}).

Fig. \ref{Schematic_rat}a,b illustrates the experimental setup and behavioral data during a 25-second segment in which the rat repeatedly traversed the linear track. Neural activity recorded during the first, middle, and last 20 cm portions of the track was selected as the windows of interest and color-coded accordingly, as shown in Fig. \ref{Schematic_rat}b. The data were then divided into 132 training and 32 test segments.
Fig. \ref{Schematic_rat}c presents the embedded neural trajectories during traversal of the three track segments using CEBRA-Time and CEBRA-Behavior. When trained on the training data and evaluated on the reserved test set, both CEBRA variants exhibit clear separation between the neural trajectories in the train trials. However, this distinction diminishes in the test trials, where the embedded time trajectories become more entangled and overlapping.
Figure \ref{Schematic_rat}d visualizes the lower-dimensional representation of the global state node, as the average embedding of the entire neuronal network using \TV{}. Train embeddings are depicted as diamonds, while test embeddings are shown as stars. The strong alignment between train and test embeddings corresponding to the same segment of the track demonstrates the high classification accuracy of \TV{} in distinguishing neural recordings from the three track portions (0–0.2 m, 0.7–0.9 m, and 1.2–1.6 m).
Fig. \ref{Schematic_rat}e depicts the train and test accuracy across training epochs using \TV{}. The training accuracy (solid blue line) and test accuracy (dashed magenta line) converge to high performance, indicating effective generalization. Finally, Fig. \ref{Schematic_rat}f presents the performance metrics (accuracy, recall, precision, and F1 score) across different embedding dimensions for \TV{}, showing that dimensions 8 and 16 yield the highest classification accuracy in the downstream task.

Having established the effectiveness of \TV{} in classifying neural activity across different track segments, we next sought to investigate the contributions of its key architectural components. To this end, we conducted an ablation study, testing four variations of our model to assess the impact of each structural element on downstream performance.

The results in Table \ref{tab:ablation_study} outline that removing Temporal Attention, replacing the Spatial-aware GRU with a conventional GRU, or replacing the Variational Graph Autoencoder with a simpler Graph Autoencoder all lead to significant performance drops across all evaluation metrics for \TV{}.

\begin{table}[h]
\centering
\caption{Ablation study of the proposed \TV{} framework.}
\label{tab:ablation_study}
\begin{tabularx}{\textwidth}{l|X|X|X|X}  
\toprule
Model Specification                                                  & Accuracy (\%) & Recall (\%) & Precision (\%) & F1-Score (\%)  \\ \hline
 Graph Autoencoder + Conventional GRU                                    & 74.12 ± 10.26          & 86.39 ± 12.92         & 75.93 ± 19.95           & 77.72 ± 6.72                \\ 
Graph Autoencoder + Spatial-aware GRU                                   & 84.71 ± 12.11           & 86.39 ± 10.84          & 84.59 ± 15.26           & 85.47 ± 10.71               \\  
Graph Autoencoder + Spatial-aware GRU + Temporal Attention            & 87.06 ± 12.00           & 91.73 ± 8.31         & 87.22 ± 15.28            & 88.10 ± 10.06                \\ 
Variational Graph Autoencoder + Spatial-aware GRU                       & 88.24 ± 4.32         & 90.83 ± 5.41          & 87.78 ± 9.63           & 88.72 ± 7.72                 \\ 
Variational Graph Autoencoder + Spatial-aware GRU + Temporal Attention &  \textbf{99.84 ± 0015}          & \textbf{99.84 ± 0015}        & \textbf{99.86 ± 0013}          & \textbf{99.84 ± 0014}      \\ \bottomrule
\end{tabularx}
\end{table}

\subsection{Primate somatosensory cortex dataset}
\begin{figure}
  \centering
  \includegraphics[width = 1\textwidth]{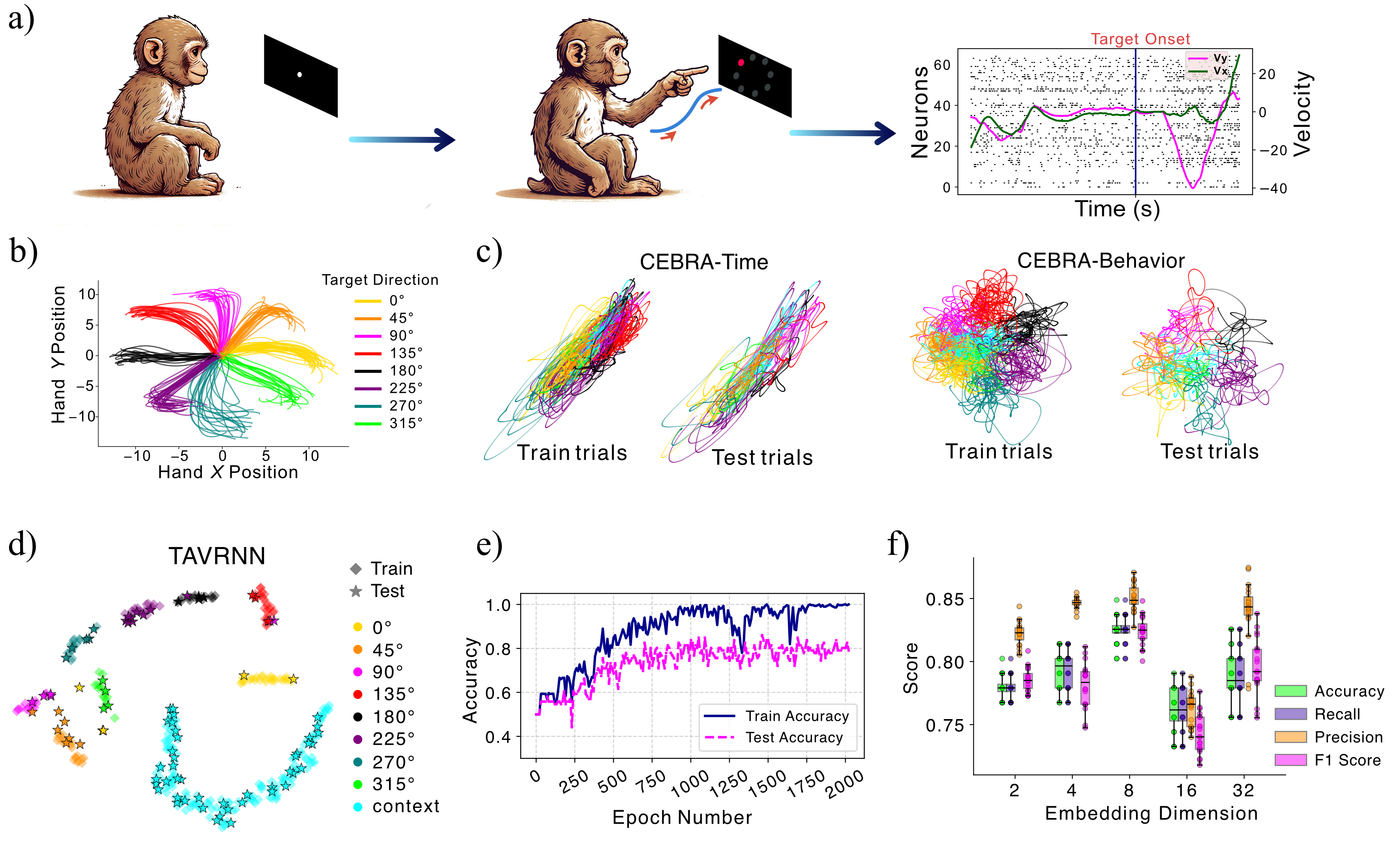}  
  \caption{\small{\textbf{a)} Schematic of primate somatosensory cortex data collection and sample neuronal activity recorded during a single trial of a target-reaching task. Each trial is divided into the context window (before target onset) and the movement phase, during which the monkey moves its hand toward the target. The hand velocity along the x-axis ($V_x$) and y-axis ($V_y$) is shown for a sample trial, illustrating the behavioral dynamics. (b) Averaged trajectories of the monkey's hand position across all trials are displayed for each of the 8 possible target directions, with different colors representing each direction.
  \textbf{c)} Embedded trajectories of all 300 train trials and 86 test trials at the population level, during both context and reaching windows for all 8 target directions, using CEBRA-Time and CEBRA-Behavior. While CEBRA-Behavior demonstrates some level of trajectory separation for the 9 classes in both train and test trials, CEBRA-Time lacks distinction between time windows with different behavioral labels when examining only population-level embeddings (trajectories over time).
  \textbf{d)} Low-dimensional representation of neuronal activity during time windows corresponding to the 9 classes of context and 8 target directions in the monkey dataset, using \TV{}. Train trials (diamonds) and test trials (stars) largely overlap. In each time window, all single neurons are embedded into a lower-dimensional space using \TV{}, and the network’s history node—representing the average activity across all neurons—is plotted for all 300 train trials and 86 test trials. \textbf{e)} Train and test set accuracy across training epochs using the \TV{} method. The train accuracy (solid blue line) and test accuracy (dashed magenta line) demonstrate convergence, with high performance achieved for both sets after sufficient training epochs. \textbf{f)} Performance metrics (accuracy, recall, precision, and F1 score) across different embedding dimensions for the \TV{} method. Each metric is represented by a distinct color, with individual points indicating the scores from 20 independent runs of the model. The embedding dimension was set to 8 for both \TV{} and CEBRA variants, with 2D visualizations used in this figure for all results.}
   }
  \label{Schematic_monkey}
\end{figure}

To further evaluate \TV{}'s ability to classify neural activity associated with complex, goal-directed behavior, we applied it to a primate somatosensory cortex dataset recorded during a visually guided target-reaching task. We used the dataset from \cite{chowdhury2020area}, which consists of neural recordings from Rhesus macaques implanted with Utah multi-electrode arrays (Blackrock Microsystems) in the arm representation of Brodmann’s area 2 of S1. The dataset captures neural activity from 65 units during an eight-direction "center-out" reaching task, where the monkey made active reaching movements (Fig. \ref{Schematic_monkey}a). Simultaneously, the monkey's hand movement was recorded across all trials with different target directions (Fig. \ref{Schematic_monkey}b).

Unlike the rat dataset, which involved continuous locomotion and three broad behavioral labels, this dataset presents a more complex classification challenge. It features discrete motor actions, fine-grained behavioral states, and a higher-dimensional behavioral space. This allows us to assess \TV{}'s ability not only to differentiate motor actions (reaching toward different targets) but also to distinguish between distinct cognitive states within a trial, such as pre-target anticipation versus movement execution. The increased complexity of this task—with nine distinct classes, including eight target directions and a pre-target context window—provides a rigorous test of \TV{}'s capacity to learn and generalize from neural data.

In this experiment, the monkey fixated at the center of the screen (context window) before reaching toward a target appearing in one of eight possible directions, while neural activity from area 2 of S1 was recorded. Each of the 193 experimental trials was divided into a context window (-100 ms before target onset) and a movement window (500 ms after target onset). The data was binned at 1 ms intervals and convolved with a Gaussian kernel ($\sigma$ = 40 ms).  
\TV{} was then applied in a downstream classification task to distinguish between nine classes: \textit{context}, \textit{0°}, \textit{45°}, \textit{90°}, $\dots$, \textit{315°} (see Fig. \ref{Schematic_monkey}d). Each trial contributed two labeled windows—context and movement—resulting in 386 labeled time windows ($2 \times 193$ trials). Of these, 300 windows from 150 trials were used for training, with the remaining windows reserved for testing (see Fig. \ref{Schematic_monkey}b). 
As the last step, within each window, we constructed a network adjacency matrix representing functional connectivity using zero-lag Pearson correlations as edges and 65 neurons as nodes. We employed graph kernels for selecting the connectivity inference method, i.e. Pearson correlation (see Supplementary information \ref{conn_inference}).

Fig. \ref{Schematic_monkey}a,b illustrates the experimental setup and behavioral data, where \ref{Schematic_monkey}b depicts the monkey’s hand positions in X and Y coordinates across all trials, color-coded based on target direction. The same color coding is applied in \ref{Schematic_monkey}c,d, where an additional class (color) represents context windows across all trials.
Fig. \ref{Schematic_monkey}c presents the embedded neural trajectories during the context and movement phases using CEBRA-Time and CEBRA-Behavior. CEBRA-Behavior shows better trajectory separation in the training data compared to CEBRA-Time. However, this separation diminishes in the test set for both methods, with CEBRA-Time showing minimal separation and significant overlap in both train and test trajectories.
Figure \ref{Schematic_monkey}d visualizes the lower-dimensional representation of the global state node, as the average embedding of the entire neuronal network using \TV{}. Train embeddings are depicted as diamonds, while test embeddings are shown as stars. The strong alignment between train and test embeddings for the same behavioral state demonstrates \TV{}'s ability to accurately classify neural activity across both context and movement windows.
Fig. \ref{Schematic_monkey}e depicts the training and test accuracy across epochs using \TV{}, showing convergence to high performance and effective generalization. Finally, Fig. \ref{Schematic_monkey}f presents classification performance metrics (accuracy, recall, precision, and F1 score) across different embedding dimensions, indicating that dimension 8 yields the best results in the downstream task.

\subsection{\DB cell culture dataset}
Having demonstrated \TV{}’s effectiveness in supervised classification tasks with clearly defined behavioral labels, we sought to explore its potential in a more challenging, unsupervised setting. Specifically, we next applied it to the \DB{} dataset, where no predefined behavioral labels exist. 
In this setting, our goal was not classification, but rather uncovering latent mechanisms that distinguish successful from unsuccessful learning, providing novel insights into how biological neuronal networks adapt in real-time environments.

Within the \DB framework, integrated in real-time with the MaxOne MEA software (Maxwell Biosystems, AG, Switzerland), \iv neuronal networks are intricately combined with \textit{in silico} computing via high-density multi-electrode arrays (HD-MEAs). Through real-time closed-loop structured stimulation and recording, these biological neuronal networks (BNNs) are then embedded in a simplified `Pong' game and showcase self-organized adaptive electrophysiological dynamics \cite{kagan2022vitro}.

\begin{figure}
  \centering
  \includegraphics[width = 1\textwidth]{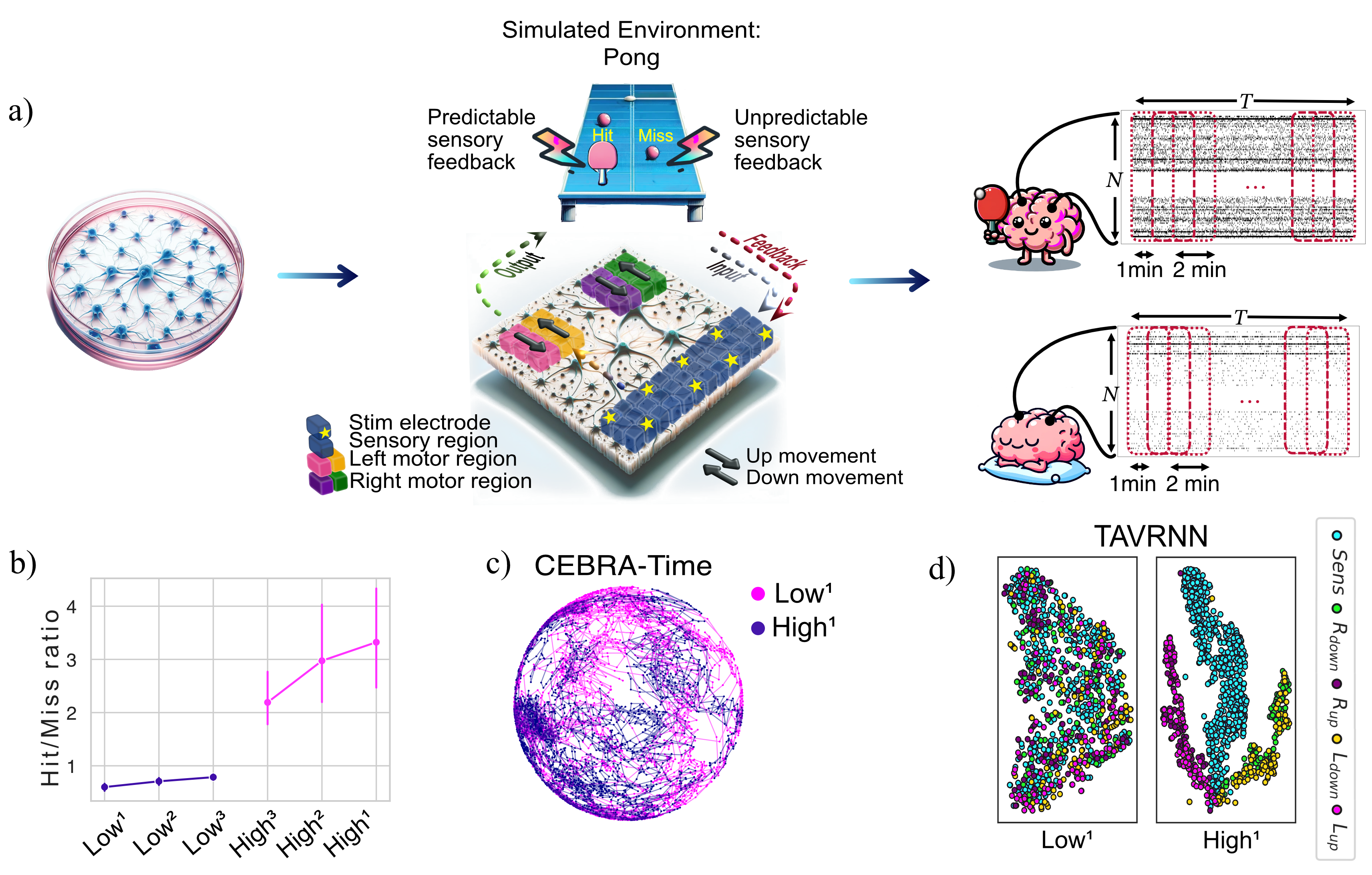}  
  \caption{\small{\textbf{a)} Schematic illustration of the \DB feedback loop, game environment, and electrode configurations.  Sample Gameplay and Rest session spike rasterplots are shown from $N = 900$ electrodes. \textbf{b)} Hit/miss ratio for the three top and bottom performing windows during gameplay averaged over all cultures. \textbf{c)} CEBRA-Time embedded trajectories for the best-performing ($High^1$) and worst-performing ($Low^1$) windows of a sample recording, illustrating that these trajectories are not distinguishable using the embedded trajectories at the population level. \textbf{d)} Lower-dimensional representation of the neuronal data for a sample culture for the best ($High^1$) and worst ($Low^1$) performing windows using \TV{}. showing all the single recorded electrodes in the embedding space. The embedding dimension was set to 8 for both \TV{} and CEBRA variants, with 2D visualizations used in this figure for all results.} }
  \label{Schematic_DB}
  
\end{figure}

Neuronal activity from 24 cultures across 437 sessions (262 'Gameplay', 175 'Rest') was recorded at 20 kHz using an HD-MEA with 900 channels. During Gameplay, sensory stimulation was delivered via 8 electrodes using rate coding (4Hz–40Hz) for the ball's $x$-axis and place coding for the $y$-axis. Paddle movement was controlled by the level of electrophysiological activity in counterbalanced "motor areas" (Fig. \ref{Schematic_DB}a). In the "motor regions," activity in half of each subregion moved the paddle "up" ($L_{up}$, $R_{up}$) and the other half moved it "down" ($L_{down}$, $R_{down}$) (Fig. \ref{Schematic_DB}a). Cultures received feedback via the same sensory regions, such that unpredictable 150 mV stimulations at 5 Hz were introduced when they missed the ball as random external inputs into the system. This was applied to arbitrary locations among the 8 sensory electrodes, at varied intervals lasting up to 4 seconds. A configurable 4-second rest period ensued before the next rally commenced. During Rest sessions, activity was recorded to move the paddle without stimulation or feedback, while outcomes were still recorded. Gameplay and Rest sessions lasted 20 and 10 minutes, respectively, with spiking events from all channels extracted in each session.\\
For each of the 24 neuronal cultures in the \DB system, spiking activity from all Gameplay and Rest trials was down-sampled from a sampling frequency of 20KHz by applying a binary OR operation within 50 ms time bins. A value of 1 was assigned if a spike occurred in any trial within the bin, and 0 otherwise. This process produced 24 binary spiking time series (one per culture), each with 900 channels, and 24,000 time points during Gameplay and 12,000 during Rest.
To investigate the single-unit interactions and dynamics of the underlying neuronal networks and their variations in game performance, we then segmented each Gameplay or Rest session into sliding windows of 2 minutes, each overlapping by half a window (i.e., 1 minute). This method generated 19 snapshots during Gameplay and 9 during Rest sessions. The selected window size ensured that the covariance matrices were not ill-conditioned based on Marchenko-Pastur distribution from random matrix theory \cite{marchenko1967distribution}. As the last step, within each window, we constructed a network adjacency matrix representing functional connectivity using zero-lag Pearson correlations as edges and 900 channels as nodes. We employed graph kernels for selecting the connectivity inference method (Pearson correlation) and determining the cutoff threshold for the \DB dataset (see Supplementary information \ref{conn_inference}).\\
Behavioral data was collected by measuring the cultures' ability to intercept the ball, quantified by the number of ‘hits’. Each rally ended with a ‘miss’, resetting the ball to a random position for a new episode.
The hit/miss ratio was defined as the ratio of accurate hits to the number of missed balls (i.e. number of rallies played) .
We computed the hit/miss ratio for each time window by averaging results across all trials for each culture. The three time windows with the highest and lowest hit/miss ratios were classified respectively as the best ($High^{1,2,3}$) and worst ($Low^{1,2,3}$) performing windows. $High^{1,2,3}$ were chosen for the main comparative analyses in the following sections. Fig. \ref{Schematic_DB}b shows the average performance levels in these six time windows. This dataset was used in a downstream clustering task with regions applied as labels to observe how channels clustered at different performance levels (see for example Fig. \ref{Schematic_DB}d). \\
Further details on this system and additional results are provided in Supplementary information \ref{cell_culture}, \ref{MEA}, \ref{DB_Config}, \ref{add_results}.

Note that, unlike conventional supervised approaches that rely on predefined behavioral labels, our goal was to allow patterns to emerge naturally rather than constrain the analysis to known performance levels. The key question in \DB{} is not whether performance varies, but rather what differences in neural activity, connectivity, and network organization drive these variations. In this setting, the unsupervised variant of CEBRA (CEBRA-Time) \cite{schneider2023learnable} failed to reveal meaningful latent structure or distinguish time trajectories corresponding to different performance levels (Fig. \ref{Schematic_DB}c). In contrast, \TV{} identified distinct patterns in the embedding space corresponding to periods of successful versus unsuccessful game performance, revealing a reorganization of recorded neuronal channels based on their functional roles within the game environment (Fig. \ref{Schematic_DB}d).

As opposed to previous methods that embed global population activity into a lower-dimensional trajectory over time, \TV{} instead learns embeddings for individual neuronal units within each behavioral time window. This finer granularity enables more precise mapping of neural activity to behavior, moving beyond whole-network trajectories to provide deeper insights into the dynamics of learning and adaptation.

\begin{figure}[h]
  \centering
  \includegraphics[width = 1\textwidth]{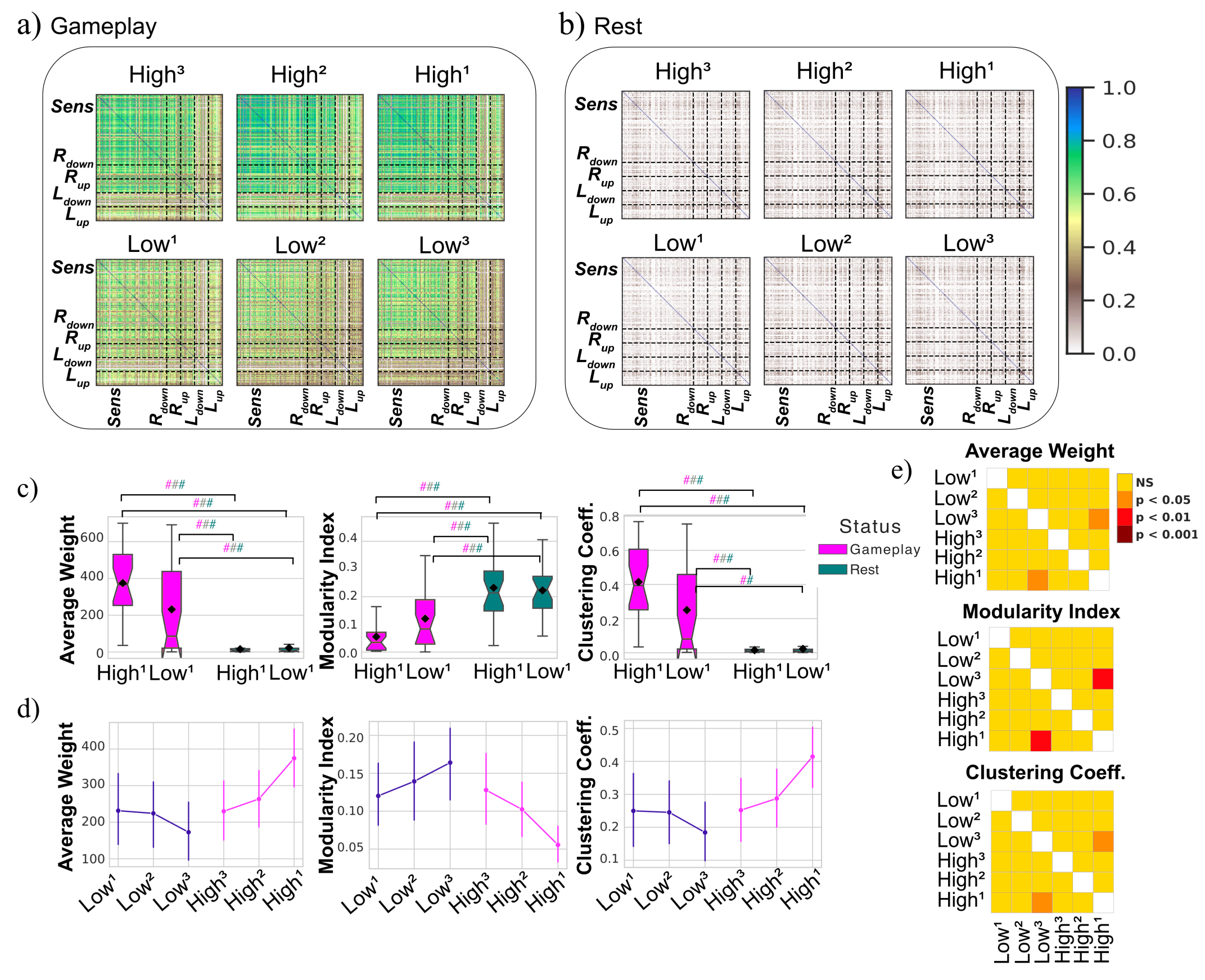}  
  \caption{Functional connectivity networks for $High^{1,2,3}$ and $Low^{1,2,3}$ windows in aggregated trials of \textbf{a)} Gameplay and \textbf{b)} Rest for a sample culture. Average weight, modularity index, and clustering coefficient for \textbf{c)} $High^{1}$ and $Low^{1}$ across all sessions. Error bands = 1 SE. $\#\#\# p < 0.001$, $\#\# p < 0.01$. \textbf{d)} Same metrics for $High^{1,2,3}$ and $Low^{1,2,3}$ in Gameplay across all recordings. Error bars = 95\% confidence intervals. \textbf{e)} Pairwise Games-Howell post-hoc test between groups.}
  \label{Net_res}
\end{figure}

Fig. \ref{Net_res}a-b depicts the connectivity networks for the top and bottom three time windows (ranked by hit/miss ratio) across all trials in a sample culture from the \DB dataset, shown separately for Gameplay and Rest conditions. The heatmaps display pairwise Pearson correlations between channels for each window. The nodes in these heatmaps are sorted by channel type on the HD-MEA, belonging to $Sens$, $L_{up}$, $R_{up}$, $L_{down}$, or $R_{down}$ regions. Across all recorded cultures, Gameplay sessions showed higher average weight, lower modularity, and lower clustering coefficients compared to Rest. Fig. \ref{Net_res}c compares these metrics for the best and worst time windows in both Gameplay and Rest, revealing significant differences between the two states. However, the absence of a significant difference between $High^{1}$ and $Low^{1}$ during Gameplay underscores the need for further investigation into the underlying mechanisms that drive successful versus unsuccessful learning, further motivating the development of \TV{}. Fig. \ref{Net_res}a-b depicts the connectivity networks for the top and bottom three time windows (ranked by hit/miss ratio) across all trials in a sample culture from the \DB dataset, shown separately for Gameplay and Rest conditions.

Thereby, to evaluate \TV{}’s effectiveness on the \DB dataset, we compared it against alternative unsupervised node-level embedding methods. Given the high dimensionality and intricate single-unit dynamics of this dataset, this comparison was crucial for assessing \TV{}’s ability to extract meaningful structure from unlabeled neuronal populations. \TV{} consistently outperformed all baselines, demonstrating its strength in uncovering the structure-function relationships that drive adaptive learning in biological neuronal systems.

\subsubsection{Baseline methods}
We used the following unsupervised node-level embedding methods as baselines, as this study focuses on drawing insights from unlabeled node sets within the \DB dataset:
\begin{itemize}
   \item \textbf{VGAE} \cite{kipf2016variational}: Unsupervised framework using a variational auto-encoder with a graph convolutional network encoder and an inner product decoder.
   \item \textbf{DynGEM} \cite{goyal2018dyngem}: Uses deep auto-encoders to generate node embeddings at each time snapshot $t$, initialized from the embedding at $t-1$.
   \item \textbf{DynAE} \cite{goyal2020dyngraph2vec}: Autoencoder model using multiple fully connected layers for both encoder and decoder to capture highly non-linear interactions between nodes at each time step and across multiple time steps.
   \item \textbf{DynRNN} \cite{goyal2020dyngraph2vec}: RNN-based model using LSTM networks as both encoder and decoder to capture long-term dependencies in dynamic graphs.
   \item \textbf{DynAERNN} \cite{goyal2020dyngraph2vec}: Employs a fully connected encoder to acquire low-dimensional hidden representations, passed through an LSTM network and a fully connected decoder.
   \item \textbf{GraphERT} \cite{beladev2023graphert}: Leverages Graph Embedding Representation using Transformers with a masked language model on sequences of graph random walks.   
\end{itemize}

\begin{figure}
  \centering
  \includegraphics[width = 0.92\textwidth]{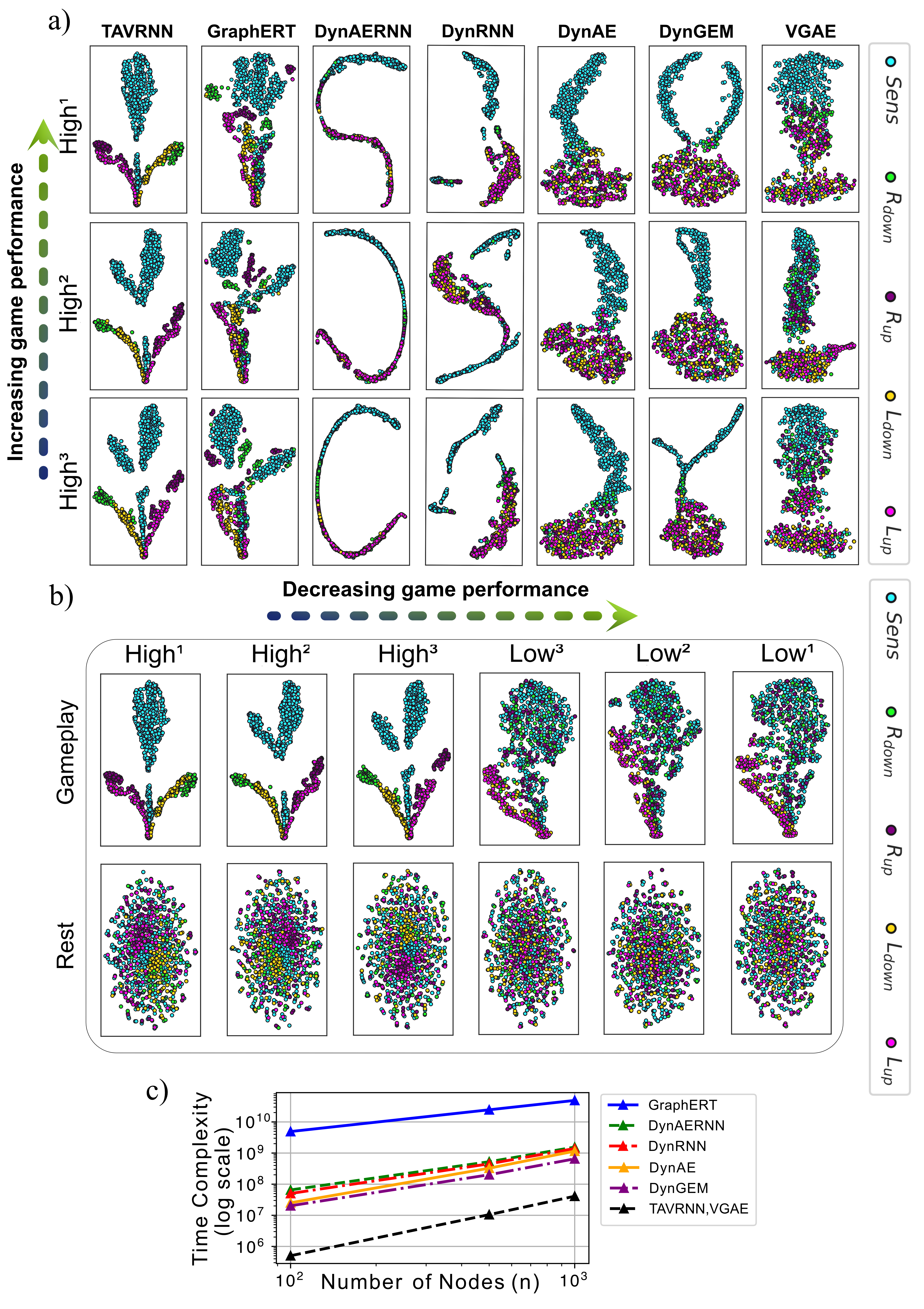}  
  \caption{\textbf{a)} t-SNE visualization of the channels in the embedding space for $High^{1,2,3}$ windows of Gameplay using \TV{} and all baseline methods for aggregated trials of a sample culture. \textbf{b)} t-SNE visualization of the channels in the embedding space using \TV{} during the top and bottom three windows ($High^{1,2,3}$ and $Low^{1,2,3}$) in terms of hit-miss-ratio during Gameplay and Rest for aggregated trials of a sample culture. Each channel is color-coded based on the predefined subregion it belongs to as shown in Fig. \ref{Schematic_DB}a. The embedding dimension was set to 8 for all methods, with 2D visualizations used in this figure. Results from additional cultures are represented in Supplementary information \ref{add_results}. \textbf{c)} Log-log plot of time complexity for all methods.}
  \label{TV_embeddings}
\end{figure}

Fig. \ref{TV_embeddings}a visualizes the embeddings for the same sample networks from Fig. \ref{Net_res} using all methods. Nodes are color-coded by their subregions on the HD-MEA. \TV{} reveals that during high game performance, nodes from different subregions (e.g., $Sens$ or motor subregions for $Up$ and $Down$ movements) form distinct clusters. The clusters become increasingly distinct as game performance reaches its highest level ($High^1$). Notably, the $Sens$ cluster overlaps with motor clusters even at peak performance, suggesting co-activation of a subgroup of $Sens$ cluster with each motor region. This clustering was not detected in the functional connectivity networks of the spiking activity (see for example Fig. \ref{Net_res}) but does accord with previous electrophysiological analysis \cite{kagan2022vitro}.
\TV{} outperforms the other baselines in separating the clusters based on the corresponding channel's subregion label. The superior performance of \TV{} compared can be linked to its capability to incorporate the temporal history of network activity. Additionally, the attention layer in the \TV{} framework enhances its effectiveness. This layer assesses the relevance of historical network activities by comparing their functional connectivity with the current snapshot, thereby significantly influencing the representation in the embedding space and leading to improved performance over the rest.
This demonstrates that successful adaptive learning requires synchronous activity between subregions, even as the modularity index of functional connectivity networks decreases during better performance. Our findings uncover the latent topology of the temporal networks revealing that clustering of subregions during successful behavior, as seen in the embedding space, highlights functional modules co-activated during optimal performance, which are not necessarily spatially proximate (see Fig. \ref{Schematic_DB}a). \\
Additionally, Fig. \ref{TV_embeddings}b represents visualization of the learned representations of the three best and three worst windows based on hit/miss ratios ($High^{1,2,3}$ and $Low^{1,2,3}$) during Gameplay and Rest, as modeled by \TV{} for all aggregated trials of the same sample culture. These visualizations reveal an absence of distinguishable clusters during the Rest state or during low-performing periods of Gameplay. However, as we progress to time windows associated with higher performance levels in the game, distinct clustering patterns emerge. 
Absence of such clustering during poor performance or Rest (as in Fig. \ref{TV_embeddings}b) implies a disruption in the coordinated activity of these modules suggesting that adaptive learning involves dynamic reorganization of neuronal circuits to optimize behavior.



Table \ref{cluster_scores} represents the comparison results during the best performing Gameplay session ($High^{1}$) across all cultures in terms of the Silhouette, Adjusted Rand Index (ARI), Homogeneity, and Completeness scores on the clustering task where channels are labeled based on their role ($Sens$, $Up$, or $Down$). We found that \TV{} outperforms all baseline methods on all metrics.  
The Silhouette score, which assesses the degree of separation among clusters, indicated some overlap in $High^{1}$ sessions. This suggests that a complete separation of clusters may not be optimal for the transmission of information between sensory and motor subregions, reflecting a functional co-activation required among channels within these clusters for goal-directed tasks.\\
The ARI evaluated the alignment between true and predicted labels where even \TV{} showed deviations from perfect alignment, highlighting the challenges of predefined neuron classifications in the \DB platform. This discrepancy stems from the absence of a definitive ground truth for defining motor subregions, complicating accurate neuron segregation. Notably, the \DB platform was originally designed considering various motor subregion configurations for $Up$ and $Down$ paddle movements, with the final regions selected based on optimal performance in experimental cultures \cite{kagan2022vitro}.
Our results indicate that neurons assigned specific roles based on their subregions did not always align with their expected activity patterns, emphasizing the complexity of predicting neuronal behavior in biological systems.\\
Analysis of homogeneity and completeness metrics revealed that neuronal clusters were heterogeneous, containing neurons from multiple classes rather than grouping all neurons of a class together—even during optimal performance. This suggests a distributed and nuanced representation of sensory and motor functions within the neuronal network, challenging predefined regional boundaries.

Note that \TV{} significantly outperforms all baselines in a task such as the clustering in the \DB dataset where the dynamics of individual nodes are crucial. Where single-unit activity is the focus of representation learning rather than population-level behavior, \TV{} excels by efficiently capturing the temporal latent dynamics of individual nodes in the graph. Additionally, our method exhibits robust performance across datasets with significantly different sampling frequencies, ranging from 40 Hz to 20/30 kHz for the rat and \DB/primate datasets.

The \TV{}framework provides a powerful tool for identifying optimized neuronal cluster organizations for task-specific functions in simulated environments like the \DB system, enhancing experimental design and efficacy. As Synthetic Biological Intelligence (SBI) advances, innovative methods like \TV{} are essential for analyzing neuronal activity and uncovering meaningful links between neural function and behavior \cite{kaganharnessing}.
Our findings highlight the complex interplay of neuronal activity in these intricate environments and emphasize the potential of our framework to enhance the understanding and design of future experiments in both biological and simulated systems.

\begin{table}
\hspace{-3cm}
  \caption{Clustering scores on the best ($High^{1}$) performing windows over all Gameplay sessions.}
  \label{cluster_scores}
  \centering
  \scriptsize{
\begin{tabular}{l|c|c|c|c}
\toprule
 Method & Silhouette  & ARI   & Homogeneity  & Completeness   \\ \midrule 
 VGAE   &  0.5385 ± 0.0337     & - 0.0014 ± 0.0004          & 0.0307 ± 0.0012            & 0.0218 ± 0.0006               \\  
                        DynGEM       & 0.4220 ± 0.0354          & 0.0035 ± 0.0056          & 0.0043 ± 0.0041         & 0.0044 ± 0.0041             \\  
                        DynAE          & 0.4133 ± 0.0366          & 0.0006 ± 0.0026          & 0.0022 ± 0.0019           & 0.0022 ± 0.0019                 \\ 
                        DynRNN                 & 0.5551 ± 0.0270         & 0.0168 ± 0.0143         & 0.0145 ± 0.0107          & 0.0149 ± 0.0110               \\ 
                         DynAERNN                 & 0.6051 ± 0.0121           & 0.1391 ± 0.0365          & 0.1053 ± 0.0312          & 0.1059 ± 0.0415                 \\
                         
                         GraphERT                  & 0.5513 ± 0.0400          &  0.6277 ± 0.1409          &  0.6046 ± 0.1110            & 0.6261 ± 0.0945               \\
                         \textbf{\TV{}}                  & \textbf{ 0.6505 ± 0.0215   }        & \textbf{ 0.8072 ± 0.0372 }        & \textbf{0.7076 ± 0.0357 }          & \textbf{0.7171 ± 0.0331}                 \\ \hline

\end{tabular}}
\vspace{-0.5cm}
\end{table}

\subsection{Time complexity analysis}
We further analyzed the time complexity of all baseline methods and compared them to \TV{}. Table \ref{time_complexity} provides the order of time complexity for one forward pass \exchange{}{on all the $n$ cells} for one time window in all methods. \exchange{}{In this table, $h_{\text{max}}$ stands for the maximum dimensionality of the hidden layers in different algorithms. See Supplementary information \ref{sec: time complexity} for more details on how the time complexities are computed and meaning of various symbols in the Table.} As demonstrated in Table \ref{time_complexity}, all the methods except GraphERT have similar orders of time complexities, but different constant coefficients. \exchange{and Fig. \ref{TV_embeddings}.c, \TV{} and VGAE exhibit the lowest time complexity, making them the most computationally efficient methods.}{} \exchange{It}{Fig. \ref{TV_embeddings}.c} shows the log-log plot of these time complexities against the number of nodes using \exchange{}{all the coefficients and hyper parameters as reported in the original paper} \exchange{the specific hyperparameter sets implemented}{} for each algorithm. \exchange{}{It shows that \TV{} and VGAE exhibit the lowest time complexity, making them the most computationally efficient methods.} In contrast, GraphERT shows the highest complexity, leading to a significant increase in run time as the number of nodes in the input graph grows. {This large time complexity is consistent with many constant coefficients we see for GraphERT in Table \ref{time_complexity}.}

\begin{table}[ht]
\centering
\caption{One forward pass time complexity per time window.}
\label{time_complexity}
\begin{tabular}{c|c} 
\toprule
\scriptsize{Method} & \scriptsize{Complexity} \\
\midrule
\scriptsize{VGAE} & \scriptsize{$\mathcal{O}\big( n \cdot \sum_{i=1}^{k} h_{i-1} \cdot h_i \big) \in \mathcal{O}\big(n^2 \cdot h_{\text{max}} \big) $} \\ 
\scriptsize{DynGEM} & \scriptsize{$\mathcal{O}\big( n \cdot \sum_{i=1}^{k+1} h_{i-1} \cdot h_i \big) \in \mathcal{O} \big(n^2 \cdot h_{\text{max}} \big)$}   \\ 
\scriptsize{DynAE} & \scriptsize{$\mathcal{O}\big( n \cdot \sum_{i=1}^{k+1} h_{i-1} \cdot h_i \big) \in \mathcal{O} \big(n^2 \cdot h_{\text{max}} \big)$} \\ 
\scriptsize{DynRNN} & \scriptsize{$\mathcal{O}\big( n \cdot \sum_{i=1}^{k+1} h_{i-1_{LSTM}} \cdot h_{i_{LSTM}} \big) \in \mathcal{O} \big(n^2 \cdot h_{\text{max}} \big)$} \\ 
\scriptsize{DynAERNN} & \scriptsize{$\mathcal{O}\big( n \cdot \sum_{i=1}^{k}  h_{i-1} \cdot h_i + h_{i-1_{LSTM}} \cdot h_{i_{LSTM}} \big) \in \mathcal{O} \big(n^2 \cdot h_{\text{max}} \big)$} \\ 
\scriptsize{GraphERT} & \scriptsize{$\mathcal{O}\big(  (\gamma \cdot |p|\cdot |q| \cdot H \cdot k) \cdot n\cdot L^2 \cdot h_{\text{max}}  \big) \in \mathcal{O} \big(n\cdot L^2 \cdot h_{\text{max}}  \big)$} \\ 
\scriptsize{\TV{}} & \scriptsize{$\mathcal{O}\big(n^2 \cdot h_{\text{max}} + n \cdot w \cdot h_{\text{max}} \big)$} \\       
\bottomrule
\end{tabular}
\end{table}

\section{Discussion}
\label{conclusions}
Understanding how neural populations evolve over time and how their dynamics relate to behavior is a fundamental challenge in systems neuroscience and neuro-inspired computation. While dimensionality reduction methods have provided valuable insights into latent neural trajectories, they often treat the population activity as a monolithic signal, ignoring the fine-grained interactions between individual neurons that drive behavioral outcomes \cite{schneider2023learnable}. Furthermore, many established methods do not explicitly account for the evolving network structure of neural populations, overlooking the temporal dependencies and reorganization that underpin adaptive learning. 

To address these limitations, we introduce \TV, a modeling framework that captures the time-varying neural connectivity and activity using probabilistic node embeddings and temporal attention. By jointly modeling both neuron-level dynamics and graph structure, \TV{} enables more precise inference of how behavioral adaptation emerges from underlying network reorganization. This dual representation is particularly powerful in biological systems, where dynamic self-organization—not static connectivity—is key to cognition and learning \cite{sporns2004organization, recanatesi2022predictive}.

We demonstrated the versatility and efficacy of \TV{}, across three distinct neural systems: (1) hippocampal electrophysiological data from a rat traversing a linear track, where ground truth labels correspond to spatial location, (2) primate somatosensory cortex electrophysiological recordings during an eight-direction reaching task, where neural activity is labeled according to movement direction, and 3) the \DB synthetic biological intelligence platform. In each case, \TV{}  demonstrated high classification accuracy, successfully capturing neural representations that correlate with behavior. Crucially, its ability to generalize across different recording modalities, sampling frequencies, and task structures suggests that the framework is broadly applicable beyond electrophysiology, extending to modalities such as fMRI time series, where no fundamental constraints limit its use.

Crucially, \TV{} revealed that optimal performance in DishBrain was associated with emergent clustering of sensory and motor subregions, despite decreasing modularity in global connectivity. This suggests that adaptive behavior may rely on transient coordination between spatially distributed but functionally aligned modules, an insight consistent with prior observations in both in vivo and synthetic systems \cite{kagan2022vitro}. These findings challenge traditional localization-based views of neural coding as understood for \textit{in vivo} organisms and supports that perhaps even more dynamic, distributed models of computation may arise in biological intelligence that are not constrained to physiological limits. Such findings are important when considering the broader scope of what functions or properties \textit{in vitro} systems may possess, allowing an iterative exploration with enhanced understanding of the neural dynamics that may drive an observed phenomena\cite{kagan2024embodied}. Moreover, the additional information provided by approaches such as \TV{}  could be further applied to microphysiological systems used in disease modeling, drug testing, and toxicology, as complex dynamics have already been observed in comparable assays \cite{watmuff2025drug}.

Despite these advantages, a limitation of the current framework is its reliance on undirected functional connectivity networks. Future iterations could incorporate directed networks \cite{razi2016connectedbrain} to differentiate between excitatory and inhibitory relationships, providing deeper insights into network-level interactions. Additionally, exploring link prediction as an extension of this framework could further enhance our understanding of how network structure evolves in response to learning and task performance.

Overall, as biological neural systems dynamically specialize and restructure through synaptic plasticity \cite{song2005highly, caroni2012structural}, identifying the principles that govern this reorganization remains an open frontier. By inferring task-relevant, temporally resolved representations directly from activity patterns, \TV{} contributes a powerful step toward mechanistic understanding of neural computation. These advances may help bridge the gap between biological and artificial learning systems—informing both foundational neuroscience and the development of interpretable, neuroadaptive AI.

\section{Methods}
\label{Methods}
\subsection{Temporal Network Construction}
\label{preprocessing}

The functional connectivity between nodes from each datasets was represented as edges in a matrix. For each time window \( t \), the corresponding temporal network is represented by a graph \( G_t \equiv (V, E) \), where \( v_i \in V \) represents a specific recording unit, and \( e_{ij} \in E \) denotes the connectivity edge between nodes \( v_i \) and \( v_j \). The structure of these dynamic network graphs \( G_t \) is captured in time-resolved adjacency matrices \( \mathbf{A}_t = [a_{t,ij}] \), with elements in \( \{0, 1\}^{N \times N} \), where $ N$ is the number of nodes. These matrices are generated by applying a threshold (as obtained from the graph kernels - see Supplementary information \ref{conn_inference}) to the functional connectivity matrices, retaining only the connections above that threshold based on absolute correlation values and setting the remainder to zero. Note that given this input structure, \TV{} is capable of handling temporal graphs from time windows of varying lengths as in the rat hippocampal or primate somatosensory cortex datasets in this study.
Additionally, each dynamic graph \( G_t \) includes node features \( \mathbf{X}_t = [x_{t,1}, \ldots, x_{t,N}]^\top \) in \( \mathbb{R}^{N \times D} \), where \( x_{t,i} \) corresponds to the feature vector of each node \( v_i \), calculated from the connection weights of each node and $D$ is the number of features. 

\vspace{-0.25cm}
\subsection{Temporal Attention-enhanced Variational Graph RNN (\TV{})}
\label{architecture}
\begin{figure}
  \centering
  \includegraphics[width = 0.9\textwidth]{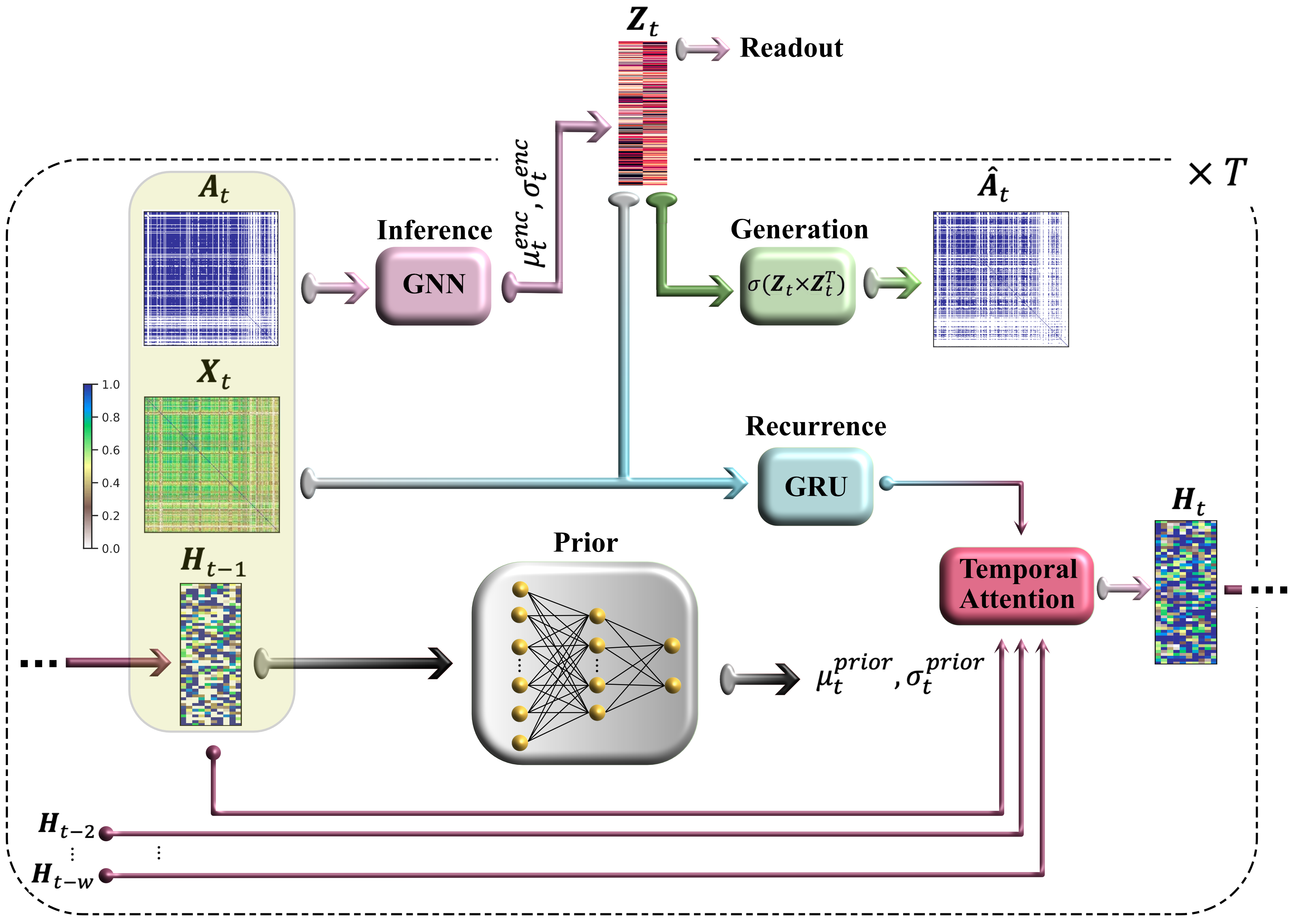}  
  \caption{\small{A schematic illustration of the \TV{} framework.}}
  \label{pipeline}
  \vspace{-0.5cm}
\end{figure}

A probabilistic \TV{} framework is developed to extract representative latent embeddings of the dynamic connectivity networks in a purely unsupervised manner. Fig. \ref{pipeline} summarises the pipeline of the introduced framework in this section. The Python implementation of our proposed framework is available at the following \href{https://github.com/TAVRNN/TAVRNN}{Github Repository}.

\vspace{-0.25cm}
\subsubsection{Spatiotemporal Variational Bayes}
We present a spatiotemporal variational Bayes objective function designed to maximize the lower bound on the log model-evidence known as the evidence lower bound (ELBO) written as $\log p_\theta(\mathcal{A}|\mathcal{X})$ or equivalently minimize its negative value known as the variational free energy (VFE) \cite{yap2023deep}. This objective is applied to a series of adjacency matrices $\mathcal{A} = \{\mathbf{A}_t\}_{t=0}^T$ from dynamic networks, based on the sequence of node features $\mathcal{X} = \{\mathbf{X}_t\}_{t=0}^T$, where $T$ is the length of the sequence. Introducing a latent embeddings sequence $\mathcal{Z} = \{\mathbf{Z}_t\}_{t=0}^T$, the VFE $\mathcal{L}^{VFE}(\theta, \phi)$ can be written via importance decomposition as:
\begin{equation}
\label{eq:1}
\mathcal{L}_{VFE}(\theta, \phi) = -\mathbb{E}_{q_\phi(\mathcal{Z}|\mathcal{X},\mathcal{A})}\left[\log \frac{p_\theta(\mathcal{A},\mathcal{Z}|\mathcal{X})}{q_\phi(\mathcal{Z}|\mathcal{X},\mathcal{A})}\right].
\end{equation}
Here, the subscripts $\theta$ and $\phi$ represent the parameters of the GNN that model the generative distribution $p_\theta(\mathcal{A},\mathcal{Z}|\mathcal{X})$ and the posterior distribution $q_\phi(\mathcal{Z}|\mathcal{X},\mathcal{A})$, respectively.
Using the following general ancestral factorizations:
\begin{equation}
\label{eq:2}
p_\theta(\mathcal{A},\mathcal{Z} \vert \mathcal{X}) = \prod_{t=0}^T p_{\theta}(\mathbf{A}_t \vert \mathbf{Z}_{\leq t}, \mathcal{X}, \mathbf{A}_{< t}) \times p_{\theta}(\mathbf{Z}_t \vert \mathcal{X}, \mathbf{A}_{< t}, \mathbf{Z}_{< t}), 
\end{equation}
\vspace{-0.5cm}
\begin{equation}
\label{eq:3}
q_\phi(\mathcal{Z} \vert \mathcal{X},\mathcal{A}) = \prod_{t=0}^T q_{\phi}(\mathbf{Z}_t \vert \mathcal{X}, \mathbf{A}_{\leq t}, \mathbf{Z}_{< t}),
\end{equation}

Eq. \ref{eq:1} is expanded to yield the sequential VFE (sVFE) as follows:
\begin{eqnarray}
\label{eq:4}
\mathcal{L}^{sVFE}(\theta, \phi) &=& -\sum_{t=0}^T \Big[\mathbb{E}_{q_{\phi}(Z_{\leq t} \vert X, A_{\le t}, Z_{<t})} \big[ \log p_{\theta}(\mathbf{A}_t \vert \mathbf{Z}_{\leq t}, \mathcal{X}, \mathbf{A}_{< t}) \big] \nonumber \\
&+& \mathcal{D}^{KL} \big[ q_{\phi}(\mathbf{Z}_t \vert \mathcal{X}, \mathbf{A}_{\leq t}, \mathbf{Z}_{< t}) \| p_{\theta}(\mathbf{Z}_t \vert \mathcal{X}, \mathbf{A}_{< t}, \mathbf{Z}_{< t}) \big] \Big].
\end{eqnarray}

Here, $\mathbf{A}_{\leq t}$ and $\mathbf{A}_{< t}$ refer to the partial sequences up to the $t^{th}$ and $(t-1)^{th}$ time samples, respectively. $\mathcal{D}^{KL}$ represents the (positive-valued) Kullback-Leibler divergence (KLD).\\
Since we want $\mathbf{Z}_t$ to represent all the information of $\mathbf{A}_t$, we replace $p_{\theta}(\mathbf{A}_t \vert \mathbf{Z}_{\le t}, \mathcal{X}, \mathbf{A}_{<t})$ in Eq. \ref{eq:4} by $p_{\theta}(\mathbf{A}_t \vert \mathbf{Z}_t)$.
Noting that Eq. \ref{eq:4} holds for any arbitrary density function $q_{\phi}$, we restrict our options to the density functions that satisfy the following equation:
\begin{equation}
    q_{\phi}(\mathbf{Z}_t \vert \mathcal{X}, \mathbf{A}_{\le t}, \mathbf{Z}_{<t}) = q_{\phi}(\mathbf{Z}_t \vert \mathbf{X}_{\le t}, \mathbf{A}_{\le t}, \mathbf{Z}_{<t})
\end{equation}
This allows us to use a simple recurrent neural network for modeling $q_{\phi}$. Also, to compute $p_{\theta}(\mathbf{Z}_t \vert \mathcal{X},\mathbf{A}_{<t},\mathbf{Z}_{<t})$ using a recurrent neural network, we simplify it by using a surrogate term $p_{\theta}(\mathbf{Z}_t \vert \mathbf{X}_{\le t},\mathbf{A}_{<t},\mathbf{Z}_{<t})$.
Applying the above substitutions into Eq. \ref{eq:4} gives:
\begin{eqnarray}
\label{eq:simpleVFE}
\mathcal{L}^{sVFE}(\theta, \phi) &=& -\sum_{t=0}^T \Big[\mathbb{E}_{q_{\phi}(Z_{\leq t} \vert X_{\le t}, A_{\le t}, Z_{<t})} \big[ \log p_{\theta}(\mathbf{A}_t \vert \mathbf{Z}_t) \big] \nonumber \\
&+& \mathcal{D}^{KL} \big[ q_{\phi}(\mathbf{Z}_t \vert \mathbf{X}_{\le t}, \mathbf{A}_{\leq t}, \mathbf{Z}_{< t}) \| p_{\theta}(\mathbf{Z}_t \vert \mathbf{X}_{\le t}, \mathbf{A}_{< t}, \mathbf{Z}_{< t}) \big] \Big].
\end{eqnarray}

The conditional probabilities in Eq. \ref{eq:simpleVFE} capture the inherent causal structure and temporal coherence of the temporal spiking activity networks. This sVFE underpins the \TV{} framework.

\subsubsection{Recurrent Graph Neural Network}
Here, we describe a model parameterization using a graph RNN for the sVFE Eq. \ref{eq:simpleVFE}. Initially, the conditional latent prior and approximate posterior in Eq. \ref{eq:simpleVFE} are assumed to follow Gaussian distributions:
\begin{subequations}
\label{eq:main5}
\begin{align}
p_{\theta}(\mathbf{Z}_t | \mathbf{X}_{<t}, \mathbf{A}_{<t}, \mathbf{Z}_{<t}) &= \mathcal{N}(\boldsymbol{\mu}_{t}^{\text{prior}}, \mathbf{\Sigma}_{t}^{\text{prior}})   \label{eq:5a} \\
q_{\phi}(\mathbf{Z}_t | \mathbf{X}_{\leq t}, \mathbf{A}_{\leq t}, \mathbf{Z}_{< t}) &= \mathcal{N}(\boldsymbol{\mu}_{t}^{\text{enc}}, \mathbf{\Sigma}_{t}^{\text{enc}}),  \label{eq:5b}
\end{align}
\end{subequations}
with isotropic covariances $\mathbf{\Sigma}_{t}^{\text{prior}} = \text{Diag}(\sigma_{t}^{\text{prior}^2}), \mathbf{\Sigma}_{t}^{\text{enc}} = \text{Diag}(\sigma_{t}^{\text{enc}^2})$, and $\text{Diag}(\cdot)$ denoting the diagonal function.
To enable gradient descent optimization of the sVFE (Eq. \ref{eq:simpleVFE}), the pairs of mean and standard deviation in Eq. \ref{eq:main5} are modeled as:
\begin{subequations}
\label{eq:main6}
\begin{align}
(\boldsymbol{\mu}_{t}^{\text{prior}}, \mathbf{\Sigma}_{t}^{\text{prior}}) &= \varphi_{\theta}^{\text{prior}}(\mathbf{H}_{t-1}) \label{eq:6a} \\
(\boldsymbol{\mu}_{t}^{\text{enc}}, \mathbf{\Sigma}_{t}^{\text{enc}}) &= \Phi_{\phi}^{\text{enc}} (\varphi_{\theta}^{\text{x}}(\mathbf{X}_t), \mathbf{H}_{t-1}, \mathbf{A}_t). \label{eq:6b}
\end{align}
\end{subequations}

In this configuration, the prior model $\varphi_{\theta}^{\text{prior}}$, the measurement feature model $\varphi_{\theta}^{\text{x}}$, and the state feature model $\varphi_{\theta}^{\text{z}}$ are designed as fully connected neural networks. Meanwhile, the encoder model $\Phi_{\phi}^{\text{enc}}$ is implemented as a GNN. The memory-embedding recurrent states $H_t$ in Eq. \ref{eq:main6} are derived as follows:
\begin{equation}
\label{eq:7}
\mathbf{H}_t = \Phi_{\theta}^{\text{rnn}}(\varphi_{\theta}^{\text{x}}(\mathbf{X}_{t}), \varphi_{\theta}^{\text{z}}(\mathbf{Z}_{t}), \mathbf{H}_{t-1}, \mathbf{A}_{t}), 
\end{equation}
where the recurrent model $\Phi_{\theta}^{\text{rnn}}$ is implemented as a spatial-aware Gated Recurrent Unit (GRU). According to Eq. \ref{eq:7}, $\mathbf{H}_t$ functions as the memory embeddings for the historical path ${\mathbf{Z}{\leq t}, \mathbf{X}{<t}, \mathbf{A}{<t}}$.\\
Subsequently, the likelihood of the adjacency matrix in Eq. \ref{eq:2} is modeled as a Bernoulli distribution:

\begin{equation}
\label{eq:8}
p_{\theta}(\mathbf{A}_t | \mathbf{Z}_t) = \text{Bernoulli} (\hat{\mathbf{A}}_t), 
\end{equation}
where $\hat{\mathbf{A}}_t$ is the reconstructed adjacency matrix, derived using a matrix product followed by sigmoid activation:
\begin{equation}
\label{eq:9}
\hat{\mathbf{A}}_t = \sigma(\mathbf{Z}_t \times \mathbf{Z}_t^T).
\end{equation}

In summary, the end-to-end integration of the prior (Eq.  \ref{eq:6a}), encoder (Eq.  \ref{eq:6b}), recurrent module (Eq. \ref{eq:7}), and inner-product decoder (Eq. \ref{eq:9}) forms a probabilistic recurrent graph autoencoder. This model first constructs sequential stochastic hierarchical latent embedding spaces on $\{\mathbf{Z}_t,\mathbf{H}_t\}^T_{t=0}$ and then utilizes these embeddings to perform stochastic estimation of the adjacency matrices $\{\hat{\mathbf{A}}_t \}^T_{t=0}$. By optimizing the sVFE (Eq. \ref{eq:simpleVFE}) with respect to the model parameters $\{ \theta, \phi \}$, these embedding spaces adapt to capture a wide array of stochastic spatiotemporal variations across dynamic networks in an entirely unsupervised manner. Further details of the method are provided in Supplementary information \ref{sVFE} and \ref{hyperparams}.

\subsubsection{Temporal Attention-based Message Passing and Spatially-aware GRU}

To more accurately reflect spatiotemporal dependencies, we reparameterized the recurrent model (Eq. \ref{eq:7}) to include a spatially-aware GRU. This modification facilitates dynamic updates of the recurrent states over time.
The update gate \( S_t \), reset gate \( R_t \), and candidate activation \( \tilde{\mathbf{H}}_t \) are calculated as:
\begin{eqnarray}
\mathbf{S}_t &=& \sigma(\Phi_{xz}(\mathbf{X}, \mathbf{A}_{t}) + \Phi_{hz}(\mathbf{H}_{t-1}, \mathbf{A}_{t})) \\
R_t &=& \sigma(\Phi_{xr}(\mathbf{X}, \mathbf{A}_{t}) + \Phi_{hr}(\mathbf{H}_{t-1}, \mathbf{A}_{t})) \\
\tilde{\mathbf{H}}_t &=& \tanh(\Phi_{xh}(\mathbf{X}, \mathbf{A}_{t}) + \Phi_{hh}(R_t \odot \mathbf{H}_{t-1}, \mathbf{A}_{t})) 
\end{eqnarray}
Finally, the output of the GRU will be computed as: 
\begin{equation}
\mathbf{\hat{H}}_t = \mathbf{S}_t \odot \mathbf{H}_{t-1} + (1 - \mathbf{S}_t) \odot \tilde{\mathbf{H}}_t 
\end{equation}

These equations describe the forward pass of our spatially-aware GRU, improving its capacity to process and incorporate spatial information through time, where \( \mathbf{X} = [\varphi_x(\mathbf{X}_{t}), \varphi_z(\mathbf{Z}_{t})]^T \). Although \( \mathbf{\hat{H}}_t \) could serve as the final value for \( \mathbf{H}_t \), given the temporal nature of our graph data, we consider a global state for the entire graph at each time step. While the GRU adds memory to the states, in our GNN structure, each node's state updates based on local information from its neighbors. For this reason, we add a hypothetical node to the graph which is connected to all other nodes. The state of this node is supposed to represent the global state of the graph. According to the dynamic nature of the graph's state, we let the model compute the final value of $\mathbf{H}_t$ through an attention mechanism on $\mathbf{\hat{H}}_t$, $\mathbf{H}_{t-1}$, $\mathbf{H}_{t-2}$, $\ldots$ and $\mathbf{H}_{t-w}$ (see Fig. \ref{pipeline}). Mathematical details of this temporal attention module are presented in Supplementary information \ref{TempAtt}. Using the above equations, \( \mathbf{H}_t \) serves as memory embeddings that capture graph-structured temporal information from previous latent state sequences. This model replaces the conventional GRU's FCNNs with single-layer GNNs \(\{\Phi_{xz}, \Phi_{hz}, \Phi_{xr}, \Phi_{hr}, \Phi_{xh}, \Phi_{hh}\}\) that incorporate a message passing scheme. This adaptation enables the GRU to efficiently leverage both the spatial topologies and temporal dependencies in dynamic graph data. For details on the parameters used in the \TV{} implementations for this study, refer to Supplementary information \ref{hyperparams}.

\newpage
\backmatter

\section*{Supplementary information}
Supplementary Materials \ref{cell_culture}-\ref{add_results}, Tables \ref{table:kernel}-\ref{full:time_complexity}, Extended Data Figs. \ref{DishBrain_config}-\ref{WL-alg}.

\section*{Acknowledgements}
...

\section*{Data availability}
The primate somatosensory cortex dataset is publicly available from the Dryad Digital Repository at \url{https://datadryad.org/dataset/doi:10.5061/dryad.nk98sf7q7}. The rat hippocampus dataset was obtained from the original study available at \url{https://www.ncbi.nlm.nih.gov/pmc/articles/PMC4919122}.The processed \DB{} data used for this study are made available at the following \href{https://github.com/TAVRNN/TAVRNN}{Github Repository}.

\section*{Code availability}
The Python implementation of our proposed framework and baseline methods is available at the following \href{https://github.com/TAVRNN/TAVRNN}{Github Repository}.

\bibliographystyle{bst/sn-aps} 

\newpage


\newcounter{extendeddata}
\newcommand{\beginsupplement}{
    \renewcommand{\thepage}{S\arabic{page}} 
    \renewcommand{\thesection}{S\arabic{section}}
    \renewcommand{\thetable}{S\arabic{table}}  
    \renewcommand{\thefigure}{S\arabic{figure}}
    \setcounter{page}{1}
    \setcounter{section}{1}
    \setcounter{table}{0}
    \setcounter{figure}{0}
    \setcounter{equation}{0}
    \newcounter{SIfig}
    \renewcommand{\theSIfig}{\arabic{SIfig}}
    }

\beginsupplement

\clearpage
\section*{Extended Data Figures} \label{sec:extended_data}
~
\begin{figure}[H]
  \centering
  \includegraphics[width = 0.96\textwidth]{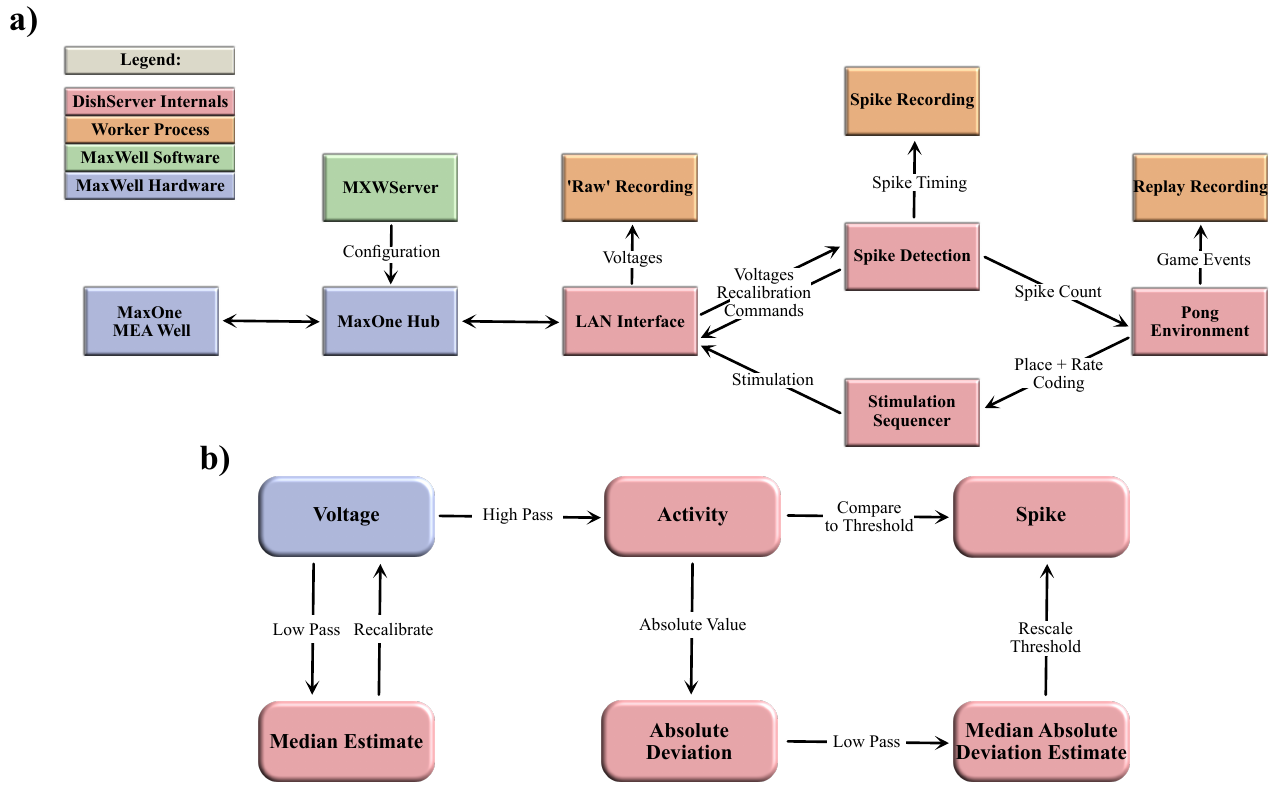}
  
  \caption{ \textbf{a, b)} Schematics of software used for DishBrain. \textbf{a)} Software components and data flow in the DishBrain closed loop system. Voltage samples flow from the MEA to the `Pong' environment, and sensory information flows from the pong environment back to the MEA, forming a closed loop. The blue rectangles mark proprietary pieces of hardware from MaxWell, including the MEA well which may contain a live culture of neurons. The green MXWServer is a piece of software provided by MaxWell which is used to configure the MEA and Hub, using a private API directly over the network. The red rectangles mark components of the `DishServer’ program, a high-performance program consisting of four components designed to run asynchronously, despite being run on a single CPU thread. The `LAN Interface’ component stores network state, for talking to the Hub, and produces arrays of voltage values for processing. Voltage values are passed to the `Spike Detection’ component, which stores feedback values and spike counts, and passes recalibration commands back to the LAN Interface. When the pong environment is ready to run, it updates the state of the paddle based on the spike counts, updates the state of the ball based on its velocity and collision conditions, and reconfigures the stimulation sequencer based on the relative position of the ball and current state of the game. The stimulation sequencer stores and updates indices and countdowns relating to the stimulations it must produce and converts these into commands each time the corresponding countdown reaches zero, which are finally passed back to the LAN Interface, to send to the MEA system, closing the loop. The procedures associated with each component are run one after the other in a simple loop control flow, but the pong environment only moves forward every 200th update, short-circuiting otherwise. Additionally, up to three worker processes are launched in parallel, depending on which parts of the system need to be recorded. They receive data from the main thread via shared memory and write it to file, allowing the main thread to continue processing data without having to hand control to the operating system and back again. \textbf{b)} Numeric operations in the real-time spike detection component of the DishBrain closed loop system, including multiple IIR filters. Running a virtual environment in a closed loop imposes strict performance requirements, and digital signal processing is the main bottleneck of this system, with close to 42 MB of data to process every second. Simple sequences of IIR digital filters is applied to incoming data, storing multiple arrays of 1024 feedback values in between each sample. First, spikes on the incoming data are detected by applying a high pass filter to determine the deviation of the activity, and comparing that to the MAD, which is itself calculated with a subsequent low pass filter. Then, a low pass filter is applied to the original data to determine whether the MEA hardware needs to be re-calibrated, affecting future samples. This system was able to keep up with the incoming data on a single thread of an Intel Core i7-8809G. Figures adapted from \cite{kagan2022vitro}. }
  \label{DishBrain_config}
\end{figure}

\begin{figure}[H]
  \centering
  \includegraphics[width = 0.95\textwidth]{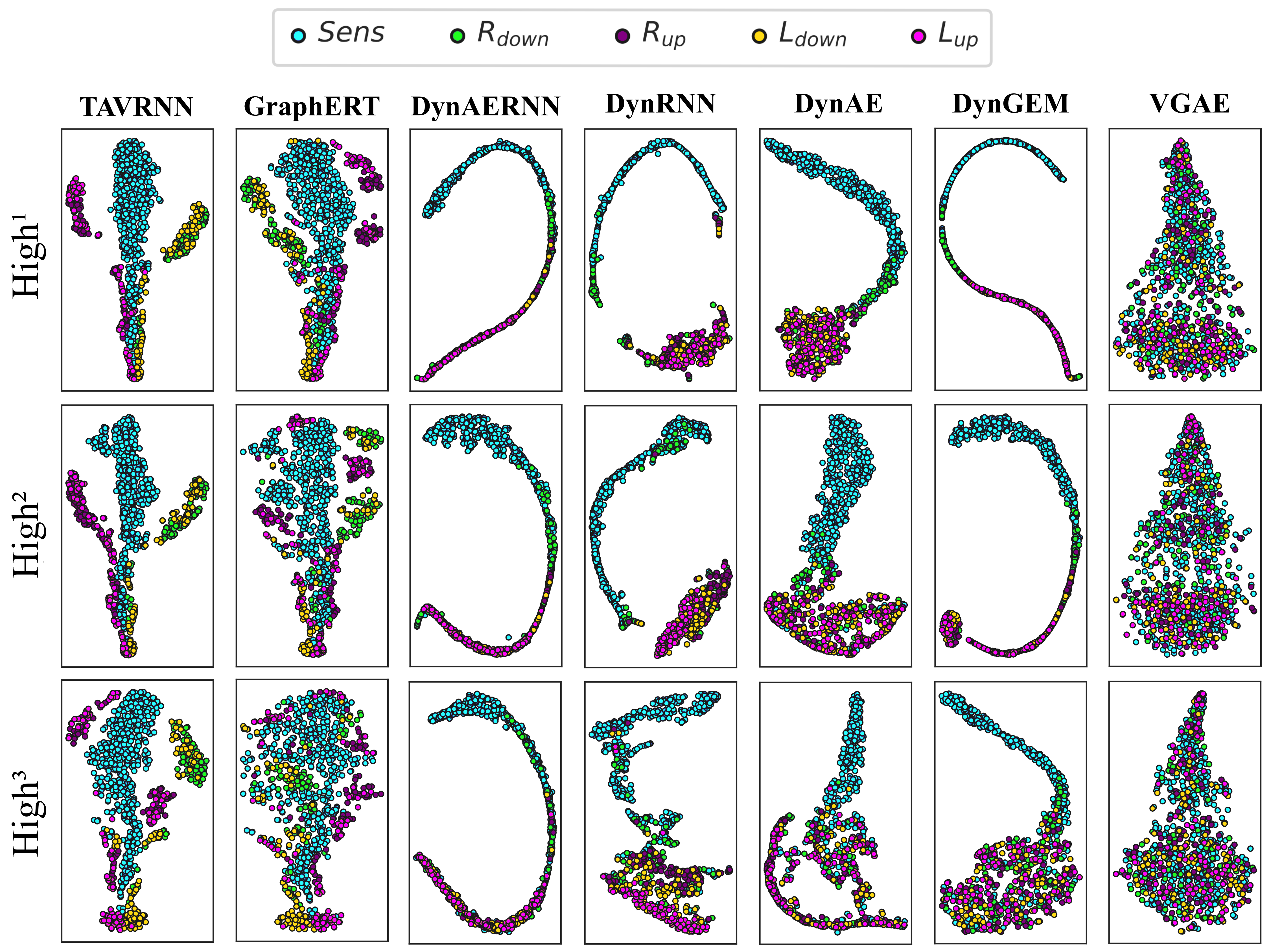}
  \caption{t-SNE visualization of the channels in the embedding space for $High^{1,2,3}$ windows of Gameplay using \TV{} and all baseline methods for aggregated trials of an additional sample culture. Each channel is color-coded based on the predefined subregion it belongs to as shown in Fig. \ref{Schematic_DB}a. }
  \label{game_embeddings_9353}
\end{figure}

\begin{figure}
  \centering
  \includegraphics[width = 0.95\textwidth]{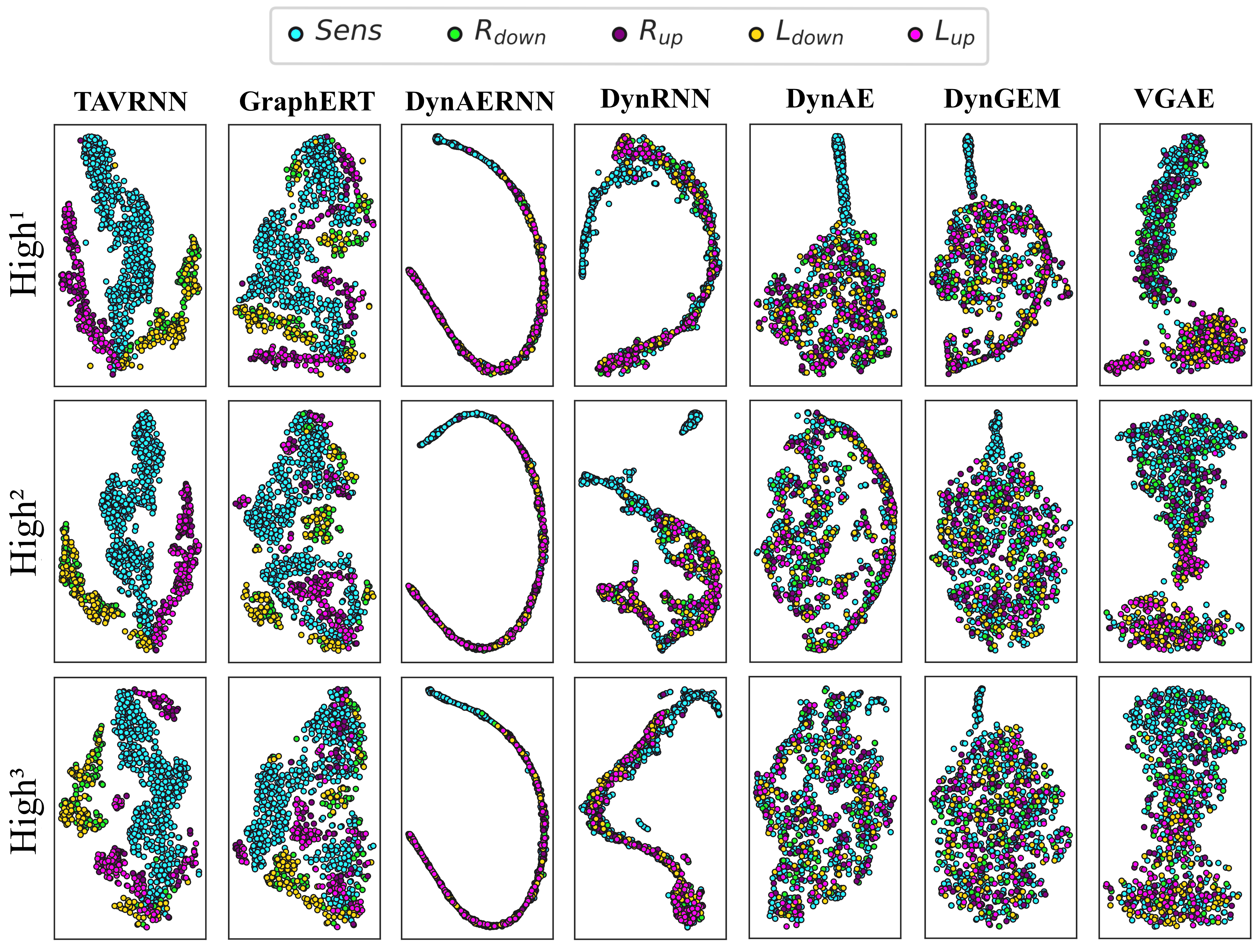}
  
  \caption{t-SNE visualization of the channels in the embedding space for $High^{1,2,3}$ windows of Gameplay using \TV{} and all baseline methods for aggregated trials of another sample culture. Each channel is color-coded based on the predefined subregion it belongs to as shown in Fig. \ref{Schematic_DB}a.}
  \label{game_embeddings_11614}
\end{figure}

\begin{figure}
  \centering
  \includegraphics[width = 0.8\textwidth]{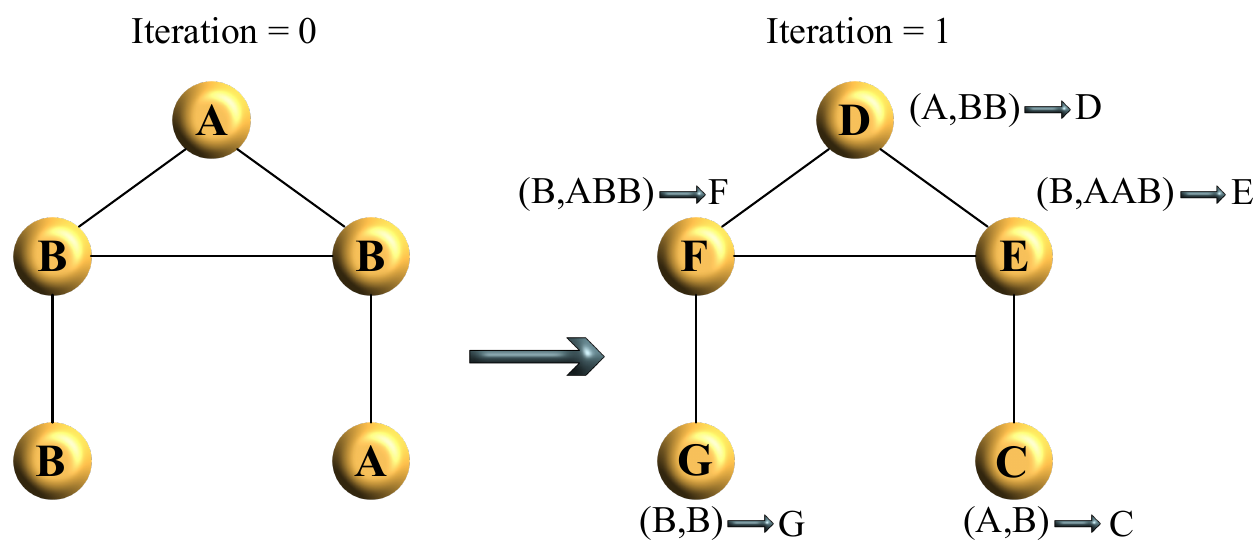}
  
  \caption{Illustration of the 1-dimensional \emph{Weisfeiler-Lehman} (1-WL) algorithm. This diagram demonstrates how the 1-WL algorithm initially encounters overlapping node labels and, through one iteration, assigns unique labels to each node based on their positions within the graph.}
  \label{WL-alg}
\end{figure}

\clearpage

\section*{Supplemental Materials}
\label{supp_mat}

\section{Cell Culture}
\label{cell_culture}
Approximately $10^6$ cells were plated on each Multielectrode Array. Neuronal cells were cultured either from the cortices of E15.5 mouse embryos or differentiated from human induced pluripotent stem cells via a dual SMAD inhibition (DSI) protocol or through a lentivirus-based NGN2 direct differentiation protocols as previously described \cite{kagan2022vitro}. Cells were cultured until plating. For primary mouse neurons, this occurred at day-in-vitro (DIV) 0, for DSI cultures this occurred at between DIV 30 - 33 depending on culture development, for NGN2 cultures this occurred at DIV 3.    
\section{MEA Setup and Plating}
\label{MEA}
MaxOne Multielectrode Arrays (MEA; Maxwell Biosystems, AG, Switzerland) was used and is a high-resolution electrophysiology platform featuring 26,000 platinum electrodes arranged over an 8 mm2. The MaxOne system is based on complementary meta-oxide-semiconductor (CMOS) technology and allows recording from up to 1024 channels. MEAs were coated with either polyethylenimine (PEI) in borate buffer for primary culture cells or Poly-D-Lysine for cells from an iPSC background before being coated with either 10 µg/ml mouse laminin or 10 µg/ml human 521 Laminin (Stemcell Technologies Australia, Melbourne, Australia) respectively to facilitate cell adhesion. Approximately $10^6$ cells were plated on MEA after preparation as per \cite{kagan2022vitro}. Cells were allowed approximately one hour to adhere to MEA surface before the well was flooded. The day after plating, cell culture media was changed for all culture types to BrainPhys™ Neuronal Medium (Stemcell Technologies Australia, Melbourne, Australia) supplemented with 1\% penicillin-streptomycin. Cultures were maintained in a low O2 incubator kept at 5\% CO2, 5\% O2, 36°C and 80\% relative humidity. Every two days, half the media from each well was removed and replaced with free media. Media changes always occurred after all recording sessions.
\section{DishBrain platform and electrode configuration}
\label{DB_Config}
The current DishBrain platform is configured as a low-latency, real-time MEA control system with on-line spike detection and recording software. The DishBrain platform provides on-line spike detection and recording configured as a low-latency, real-time MEA control. The DishBrain software runs at 20 kHz and allows recording at an incredibly fine timescale. This setup captured neuronal electrical activity and provided long-term, safe external electrical stimulation through biphasic pulses that elicited action potentials in neurons, as detailed in previous studies \cite{ruaro2005toward}.
There is the option of recording spikes in binary files, and regardless of recording, they are counted throughout 10 milliseconds (200 samples), at which point the game environment is provided with how many spikes are detected in each electrode in each predefined motor region as described below. Based on which motor region the spikes occurred in, they are interpreted as motor activity, moving the ‘paddle’ up or down in the virtual space. As the ball moves around the play area at a fixed speed and bounces off the edge of the play area and the paddle, the pong game is also updated at every 10ms interval. Once the ball hits the edge of the play area behind the paddle, one rally of pong has come to an end. The game environment will instead determine which type of feedback to apply at the end of the rally: random, silent, or none. Feedback is also provided when the ball contacts the paddle under the standard stimulus condition. A ‘stimulation sequencer’ module tracks the location of the ball relative to the paddle during each rally and encodes it as stimulation to one of eight stimulation sites. Each time a sample is received from the MEA, the stimulation sequencer is updated 20,000 times a second, and after the previous lot of MEA commands has completed, it constructs a new sequence of MEA commands based on the information it has been configured to transmit based on both place codes and rate codes. The stimulations take the form of a short square bi-phasic pulse that is a positive voltage, then a negative voltage. This pulse sequence is read and applied to the electrode by a Digital to Analog Converter (or DAC) on the MEA. A real-time interactive version of the game visualiser is available at \url{https://spikestream.corticallabs.com/}. Alternatively, cells could be recorded at ‘Rest’ in a Gameplay environment where activity was recorded to move the paddle but no stimulation was delivered, with corresponding outcomes still recorded. Using this spontaneous activity alone as a baseline, the Gameplay characteristics of a culture were determined.  Low level code for interacting with Maxwell API was written in C to minimize processing latencies-so packet processing latency was typically $<$50 $\mu$s. High-level code was written in Python, including configuration setups and general instructions for game settings. A 5 ms spike-to-stim latency was achieved, which was substantially due to MaxOne's inflexible hardware buffering. Extended Data Fig. \ref{DishBrain_config} illustrates a schematic view of Software components and data flow in the DishBrain closed loop system.

\section{Connectivity Inference Mechanisms}
\label{conn_inference}

Methods for inferring connectivity are broadly categorized into two types: \emph{model-free} and \emph{model-based} approaches. Model-free methods rely on descriptive statistics and do not presuppose any specific underlying data generation mechanism, making them versatile for initial analyses. In contrast, model-based methods involve hypothesizing a mathematical model to elucidate the underlying biological processes by estimating its parameters and structure. Typically, these methods analyze time-series data, such as spike trains from individual neurons. However, recent advances have enabled studies to integrate spike inference with connectivity analysis directly from time-series data \cite{de2018connectivity}. In this work, we focus on utilizing the model-free methods.

Model-free methods do not presuppose any specific mechanisms underlying the observed data, offering a simpler alternative to model-based approaches. However, these methods do not facilitate the generation of activity data crucial for model validation or predictive analysis. Model-free techniques are primarily divided into two categories: those employing descriptive statistics such as Pearson correlation coefficient (PC) and cross-correlation (CC) and those utilizing information-theoretic measures such as Mutual information (MI), and Transfer entropy (TE) \cite{de2018connectivity, cohen2011measuring, knox1981detection, garofalo2009evaluation,schreiber2000measuring,stetter2012model,ito2011extending}.

\subsection{Graph kernels}
\label{graph_kernels}
In light of the diversity of connectivity inference methods discussed previously, each method can generate distinct graph representations from identical datasets. To extract meaningful insights from these varied representations, it is essential to employ a comparison methodology. However, graph comparison is computationally challenging. Ideally, one would verify if two graphs are exactly identical, a problem known as graph isomorphism, which is NP-complete \cite{johnson2005np}. This complexity renders the task computationally prohibitive for large graphs.

To circumvent these difficulties, kernel methods offer a viable alternative. Kernels are functions designed to measure the similarity between pairs, enabling the transformation of objects into a high-dimensional space conducive to linear analysis methods. Graph kernels, specifically, facilitate the comparison of graphs by evaluating their structure, topology, and other attributes, thus proving instrumental in machine learning applications for graph data, such as clustering and classification \cite{hofmann2008kernel,kriege2020survey,borgwardt2020graph}.

Graph kernels vary in their approach to measuring similarity. Some rely on neighborhood aggregation, which consolidates information from adjacent nodes to form local feature vectors \cite{shervashidze2011weisfeiler,morris2019weisfeiler,neumann2016propagation}, while others utilize assignment and matching techniques to establish correspondences between nodes in different graphs \cite{frohlich2005optimal}. Additionally, some kernels identify and compare subgraph patterns \cite{shervashidze2009efficient}, and others analyze walks and paths to capture structural nuances \cite{borgwardt2005shortest}.

Here we concentrate on neighborhood aggregation methods, particularly pertinent for analyzing connectivity graphs derived from neuronal recordings, typically involving fewer than 1000 nodes without definitive node labels. These methods are also foundational for the graph neural network models. We exemplify this approach with the 1-dimensional \emph{Weisfeiler-Lehman} (1-WL) algorithm \cite{shervashidze2011weisfeiler}, illustrating its application and effectiveness.

\textbf{\emph{Weisfeiler-Lehman} Algorithm}
The \emph{Weisfeiler-Lehman} (WL) graph kernel is a sophisticated approach for computing graph similarities, which leverages an iterative relabeling scheme based on the \emph{Weisfeiler-Lehman} isomorphism test. This method extends the basic graph kernel framework by incorporating local neighborhood information into the graph representation, making it particularly effective for graph classification tasks.

Consider a graph \( G = (V, E, \ell) \), where \( V \) is the set of vertices, \( E \) is the set of edges, and \( \ell: V \rightarrow \Sigma \) is a labeling function that maps each vertex to a label from a finite alphabet \( \Sigma \). Initially, each vertex is assigned a label based on its original label or degree.

Define \( \ell^0 = \ell \). At each iteration \( i \), a new labeling \( \ell^i \) is computed as follows:
\[
\ell^{i+1}(v) = \text{HASH} \left( \ell^i(v), \{\!\!\{ \ell^i(u) \mid u \in N(v) \}\!\!\} \right)
\]
where \( N(v) \) denotes the set of neighbors of vertex \( v \) and \( \{\!\!\{ \cdot \}\!\!\} \) denotes a multiset, ensuring that the labels of neighboring vertices are considered without regard to their order. The function HASH maps the concatenated labels to a new, unique label. The algorithm continues iteratively, relabeling vertices until the labels converge or no new labels are produced (Extended Data Fig. \ref{WL-alg}).

After each iteration \( i \), compute a feature vector \( \phi^i(G) \) as the histogram of the labels across all vertices:
\[
\phi^i(G) = \left( \#\{v \in V \mid \ell^i(v) = k\} \right)_{k \in \mathcal{K}}
\]
where \( \mathcal{K} \) is the set of all possible labels at iteration \( i \).

The WL kernel between two graphs \( G \) and \( G' \) is defined as the sum of base kernel evaluations on the corresponding histograms at each iteration:
\[
K(G, G') = \sum_{i=0}^h K_{\text{base}} \left(\phi^i(G), \phi^i(G') \right)
\]
where \( K_{\text{base}} \) is typically chosen to be the linear kernel \( K_{\text{base}}(\phi, \phi') = \phi \cdot \phi' \), and \( h \) is a predefined number of iterations, determining the depth of neighborhood aggregation.

In this study, we analyzed 437 recording sessions, comprising 262 Gameplay and 175 Rest sessions, to construct functional connectivity graphs. These graphs were derived using four distinct network inference algorithms: Zero-lag Pearson Correlations (PC), Cross-Correlation (CC), Mutual Information (MI), and Transfer Entropy (TE). For the PC analysis, connectivity matrices were thresholded at varying levels \( t \in \{0, 20, 40, 60, 80\} \% \), retaining only the strongest connections as determined by their absolute correlation values. For both CC and TE, we explored delay values \( d \in \{1, 2, 3, 4\} \). Each method produced 437 distinct networks.

Subsequently, a Weisfeiler-Lehman (WL) graph kernel with depth \( h = 4 \) was utilized to compute the kernel matrix \( \mathbf{K} \), which was then employed in a Support Vector Machine (SVM) classifier to distinguish between Gameplay and Rest sessions. Classification effectiveness was evaluated through a 5-fold cross-validation on the \DB dataset, achieving the results summarized in Table \ref{table:kernel}. Notably, classification performance for CC and TE improved with increasing delay values, reflecting enhanced discriminative power of the graph kernels with longer embedding lengths. However, this increase in delay also introduced greater computational complexity, presenting challenges in scalability and traceability.

\begin{table}[ht]
\centering
\caption{Network inference method performance on \DB dataset}
\label{table:kernel}
\begin{tabular}{@{}lcc@{}}
\toprule
Network inference method & Avg. accuracy & Std. dev. \\ \midrule
PC (t = 0\%)                       & 0.672        & 0.062     \\
PC (t = 20\%)                       & 0.735         & 0.073     \\
PC (t = 40\%)                       & \textbf{0.831}         & \textbf{0.034}     \\
PC (t = 60\%)                       & 0.552         & 0.019     \\
PC (t = 80\%)                       & 0.464         & 0.047     \\

CC (d=1)       & 0.432        & 0.126     \\
CC (d=2)      & 0.546          & 0.082    \\
CC (d=3)       & 0.698        & 0.092     \\
CC (d=4)       & 0.763         & 0.103     \\

MI       & 0.722         & 0.057     \\

TE (d=1)       & 0.657         & 0.073     \\
TE (d=2)      & 0.688         & 0.112     \\
TE (d=3)       & 0.731         & 0.028     \\
TE (d=4)       & 0.794         & 0.063     \\ \bottomrule
\end{tabular}
\end{table}

\section{Marchenko-Pastur Distribution and Shuffling Procedure}
\label{marchenko}

In random matrix theory, the Marchenko-Pastur (MP) distribution describes the asymptotic behavior of the eigenvalues of large-dimensional sample covariance matrices. Consider a random matrix \( \mathbf{A} \in \mathbb{R}^{p \times n} \), where \( p \) represents the number of variables (e.g., neurons or channels) and \( n \) represents the number of observations (e.g., time points). The sample covariance matrix is defined as:

\[
\mathbf{C} = \frac{1}{n} \mathbf{A}^T \mathbf{A}
\]

As both \( p \) and \( n \) grow large, while the ratio \( \eta = \frac{p}{n} \) remains constant, the empirical distribution of the eigenvalues of \( \mathbf{C} \) converges to the Marchenko-Pastur distribution \cite{marchenko1967distribution}:

\[
\rho(\lambda) = \frac{\sqrt{(\lambda_+ - \lambda)(\lambda - \lambda_-)}}{2\pi \sigma^2 \lambda \eta}
\]

for \( \lambda \in [\lambda_-, \lambda_+] \), where $\sigma$ is the variance of the entries of matrix $\mathbf{A}$ and:

\[
\lambda_{\pm} = \sigma^2 \left(1 \pm \sqrt{\eta}\right)^2
\]

In the case where \( \eta > 1 \), which holds for our data (\( p \) is large relative to \( n \)), the MP distribution suggests that most of the eigenvalues will be close to zero. As a result, the sample covariance matrix is likely to be ill-conditioned, and hence unreliable for further analysis.

\subsection{Shuffling Procedure for Correlation Analysis}

To account for potential spurious correlations due to ill-conditioning of the sample covariance matrix, we perform a shuffling control procedure:

\begin{enumerate}
    \item \textbf{Shuffle Time Points:} The time points of each channel are independently shuffled while maintaining the channel identity. This process destroys any temporal correlation, ensuring that the correlation between channels is not influenced by the original time structure.
    \item \textbf{Multiple Iterations:} The shuffling procedure is repeated multiple times (e.g., we chose 1000 iterations) to build a null distribution of correlations for each pair of channels.
    \item \textbf{Confidence Intervals:} Based on the null distribution obtained from the shuffled data, we compute confidence intervals for each pair of channels. Correlation values from the original data that lie outside of the \( 95\%\) confidence interval are considered statistically significant.
\end{enumerate}

This approach provides a robust method for identifying significant correlations in the presence of potential ill-conditioning of the sample covariance matrix.

\section{CEBRA model setting}
\label{CEBRA_setting}
Following the approach in \cite{schneider2023learnable}, all experiments implementing the CEBRA variants used a receptive field of ten samples. The model architecture consisted of a convolutional network with five time-convolutional layers. The first layer had a kernel size of two, while the next three had a kernel size of three and incorporated skip connections. The final layer, also with a kernel size of three, projected the hidden dimensions to the output space. Gaussian error linear unit (GELU) activation functions \cite{hendrycks2016gaussian} were applied after each layer except the last. The feature vector was normalized after the final layer. An output dimension of 8 was maintained across all datasets, and all other architectural parameters were kept identical to those in Schneider et al. \cite{schneider2023learnable}.

\section{Unsupervised sequential VFE (sVFE) loss}
\label{sVFE}

In a Variational Graph Auto Encoder (VGAE), an encoder network is responsible for learning the latent embeddings $\{\mathbf{Z}_t\}_{t=0}^T$, which capture the representation of nodes in a reduced-dimensional space. The probablity of an edge between nodes $i$ and $j$ in the reconstructed graph is determined by the inner product of their respective latent embeddings, $\mathbf{Z}_{t,i}$ and $\mathbf{Z}_{t,j}$. This process is usually accompanied by a sigmoid activation function to constrain the output values between 0 and 1:

\begin{equation}
\label{eq:s1}
\hat{a}_{t,ij} = \sigma(\mathbf{Z}_{t,i} \cdot \mathbf{Z}_{t,j}^T). 
\end{equation}
In this context, $\sigma$ represents the sigmoid function, \(\mathbf{Z}_{t,i}\) refers to the \(i\)th row of the matrix \(\mathbf{Z}_t\), and $\hat{a}_{t,ij}$ corresponds to the \((i, j)\)th element of the matrix \(\hat{\mathbf{A}}_t\), indicating the predicted probability of an edge between nodes $i$ and $j$ at time $t$.

Considering that $\hat{a}_{t,ij}$ indicates the probability of an edge, the likelihood of the observed adjacency matrix $\mathbf{A}_t$ based on the embeddings can be independently modeled for each edge using a Bernoulli distribution:

\begin{equation}
\label{eq:s2}
p_{\theta}(\mathbf{A}_t | \mathbf{Z}_{\leq t}, \mathbf{X}_{< t}, \mathbf{A}_{< t}) = \prod_{i,j=1}^N \hat{a}_{t,ij}^{a_{t,ij}} (1-\hat{a}_{t,ij})^{1-a_{t,ij}}.
\end{equation}

In this case, $a_{t,ij}$ represents the actual entry in the adjacency matrix $\mathbf{A}_t$, signifying the presence, absence, or weight (for weighted graphs) of an edge between nodes $i$ and $j$.

The log-likelihood of the adjacency matrix, $\log p_{\theta}(\mathbf{A}_t | \mathbf{Z}_{\leq t}, \mathbf{X}_{< t}, \mathbf{A}_{< t})$, can be expressed as the negative binary cross entropy (BCE):

\begin{equation}
    \label{eq:s3}
    \mathcal{L}^{\text{BCE}}(\theta, \phi) = \sum_{i,j=1}^{N} \Big[ a_{t,ij} \log \hat{a}_{t,ij} + (1 - a_{t,ij}) \log (1 - \hat{a}_{t,ij}) \Big]. 
\end{equation}

We approximate the first expectation term in the sequential VFE (sVFE) using Monte Carlo integration as follows:

\begin{equation}
    \label{eq:s4}
    \mathbb{E}_{q_{\phi}(z_{t}|x_{\leq t})} \left[ \log p_{\theta}(\mathbf{A}_t|\mathbf{Z}_{\leq t}, \mathbf{X}_{<t}, \mathbf{A}_{<t}) \right] = \frac{1}{M} \sum_{k=1}^M \mathcal{L}^{\text{BCE}}(\mathbf{Z}_t^k). 
\end{equation}

Here, \(k\) represents the particle index, and \(M\) refers to the number of particles, which may be set to 1 when the mini-batch size is sufficiently large \cite{kingma2013auto}.

Latent particles \(\mathbf{Z}_t^k\) are sampled from \(q_{\phi}(\mathbf{Z}_t | \mathbf{X}_{\leq t}, \mathbf{A}_{\leq t}, \mathbf{Z}_{<t})\) as described by Eq. (\ref{eq:5b}), utilizing the reparameterization trick \(\mathbf{Z}_t^k = \mu_t^{\text{enc}} + \sigma_t^{\text{enc}} \odot \epsilon_t^k\), where \(\epsilon_t^k\) is drawn from \(\mathcal{N}(0, I)\) and \(\odot\) represents the Hadamard (element-wise) product. Recurrent state particles \(\mathbf{H}_t^k\) are derived using Eq. (\ref{eq:7}), based on \(\mathbf{Z}_{t-1}^k\) and the previous time-step's state \(\mathbf{H}_{t-1}^k\).

Additionally, an analytical solution for the Kullback-Leibler divergence \(D_{\text{KL}}\) in the sequential VFE Eq. (\ref{eq:4}) can be derived in closed form as:
\begin{equation}
\label{eq:s5}
    D_{\text{KL}}(\theta, \phi) = \frac{1}{2} \sum_{i,j=1}^{N,D} \left[ \frac{\sigma_{t,ij}^{\text{enc}2}}{\sigma_{t,ij}^{\text{prior}2}} - \log \frac{\sigma_{t,ij}^{\text{enc}2}}{\sigma_{t,ij}^{\text{prior}2}} + \frac{(\mu_{t,ij}^{\text{enc}} - \mu_{t,ij}^{\text{prior}})^2}{\sigma_{t,ij}^{\text{prior}2}} - 1 \right] 
\end{equation}

This KLD loss is deterministic, thereby eliminating the need for Monte Carlo approximation. It quantifies the statistical distance between the conditional prior as specified in Eq. (\ref{eq:5a}) and the approximate posterior in Eq. (\ref{eq:5b}). Optimizing this measure strengthens the causality within the latent space, as the prior Eq. (\ref{eq:6a}) focuses on the influence of preceding graphs and embeddings $\{ \mathbf{X}<t, \mathbf{A}<t, \mathbf{Z}<t \}$.

By integrating Eq. (\ref{eq:s4}) and Eq. (\ref{eq:s5}) into Eq. (\ref{eq:4}), we formulate an unsupervised sVFE loss that forms the foundation of the proposed \TV{} framework:

\begin{equation}
\label{eq:s6}
\begin{aligned}
    \mathcal{L}^{\text{\TV{}}}(\theta, \phi) &= \mathcal{L}^{\text{BCE}}(\theta, \phi) + \mathcal{D}^{\text{KL}}(\theta, \phi) \\
    &= \underbrace{\frac{1}{M} \sum_{t=0}^T \sum_{k=1}^M \sum_{i,j=1}^{N} \left[ a_{t,ij} \log \sigma \left( \mathbf{Z}_{t}^k \times \mathbf{Z}_{t}^{k^T} \right)  + (1 - a_{t,ij}) \log \left(1 - \sigma \left( \mathbf{Z}_{t}^k \times \mathbf{Z}_{t}^{k^T} \right) \right) \right]}_{\mathcal{L}^{\text{BCE}}(\theta, \phi)} \\
    &\quad + \underbrace{\frac{1}{2} \sum_{t=0}^T \sum_{i,j=1}^{N} \left[ \frac{{(\sigma_{t,ij}^{\text{enc}}+\epsilon})^2}{(\sigma_{t,ij}^{\text{prior}}+\epsilon)^2} - \log \frac{{(\sigma_{t,ij}^{\text{enc}}+\epsilon})^2}{(\sigma_{t,ij}^{\text{prior}}+\epsilon)^2} + \frac{(\mu_{t,ij}^{\text{enc}} - \mu_{t,ij}^{\text{prior}})^2}{(\sigma_{t,ij}^{\text{prior}}+\epsilon)^2} - 1 \right]}_{\mathcal{D}^{\text{KL}}(\theta, \phi)}.
\end{aligned}
\end{equation}

\section{Temporal attention mechanism}
\label{TempAtt}
The goal of this section is to present the mathematical details of the temporal attention mechanism for computing $\mathbf{H}_t$ for $\mathbf{\hat{H}}_t$ and $\mathbf{H}_{t-1}$, $\mathbf{H}_{t-2}$, $\ldots$ $\mathbf{H}_{t-w}$.
Let the $d_h$ dimensional row vector $\overline{s}_i$ present the global state of the graph at time step $i$. \footnote{For $i<t$, $\overline{s}_i$ is equal to that row of $\mathbf{H}_i$ which corresponds to the hypothetical node that is connected to all other nodes. Also, $\overline{s}_t$ is equal to the corresponding row of $\mathbf{\hat{H}}_t$.} Also let $\mathbf{\overline{S}}$ be a $(w+1) \times (w+1)$ matrix that its $i$-th row is equal to $\overline{s}_{t-w-1+i}$.
We compute the query vector $q$ and the key matrix $K$ as follows:
\begin{equation}
    q = \overline{s}_t \times \mathbf{W}_q + b_q 
\end{equation}
\begin{equation}
    \mathbf{K} = \mathbf{\overline{S}} \times \mathbf{W}_k + b_k  
\end{equation}
Here, the $d_h \times d_k$ matrices $\mathbf{W}_q$ and $\mathbf{W}_k$, and also the $d_k$ dimensional row vectors $b_q$ and $b_k$ are learnable parameters of our model. Then, the attention vector $\alpha$, which is a $w+1$ dimensional row vector, will be defined as:
\begin{equation}
    \alpha = \text{softmax}\left(\frac{q \times K^T}{\sqrt{d_k}}\right).  
\end{equation}

Let us define the value matrices as follows:
\begin{equation}
    \mathbf{V}_i = \mathbf{H}_{t-w-1+i} \times \mathbf{W}_v + b_v \verb|   | \forall 1 \le i \le w ~,  
\end{equation}
and 
\begin{equation}
    \mathbf{V}_{w+1} = \mathbf{\hat{H}}_t \times \mathbf{W}_v + b_v~,  
\end{equation}
where the $d_h \times d_h$ matrix $\mathbf{W}_v$ and the $d_h$ dimensional row vector $b_v$ are the other learnable parameters of our model.

Finally, the state matrix $\mathbf{H}_t$ will be computed as follows:
\begin{equation}
    \mathbf{H}_t = \displaystyle\sum_{i=1}^w \alpha_i \times \mathbf{V_i} ~.  
\end{equation}

\section{\TV{} model training hyperparameters}
\label{hyperparams}

All the experiments were run on a 2.3 GHz Quad-Core Intel Core i5. PyTorch 1.8.1 was used to build neural network blocks. 

We configured our \TV{} model by employing graph-structured GRU-Attention with a single recurrent hidden layer consisting of 32 units. The window size $w$ in the attention mechanism is set to the maximum possible for every time step, allowing the model to attend to all previous time steps, including the very first one. The functions $\varphi_{\theta}^{\text{x}}$ and $\varphi_{\theta}^{\text{z}}$ in Eqs. (\ref{eq:6b}) and (\ref{eq:7}) are implemented using a 32-dimensional fully-connected layer. For the function $\varphi_{\theta}^{\text{prior}}$ in Eq. (\ref{eq:6a}), we use two 32 and 8 dimensional fully-connected layers. To model $\boldsymbol{\mu}_{t}^{\text{enc}}$ and $\mathbf{\Sigma}_{t}^{\text{enc}}$ we employ a 2-layer GCN with 32 and 8 layers, respectively with 8 being selected as the readout dimension ($Z_t$) for the experiments in this study. Our model is initialized using Glorot initialization \cite{glorot2010understanding}. The learning rate for training is set to 0.01. Training is performed over 1000 epochs using the Adam SGD optimizer \cite{kingma2014adam}. 

Additionally, we utilize the validation set based on a 5-fold Cross-validation for the early stopping.

The implementation of our proposed model is available at the following \href{https://github.com/TAVRNN/TAVRNN}{Github Repository}.

\section{Time Complexity Analysis}
\label{sec: time complexity}

In \exchange{the following}{this} section, we will compute the time complexity for each method. \exchange{The time complexity}{This analysis} provides insights into the computational cost and efficiency of different \exchange{approaches used in graph and temporal data analysis}{methods for representation learning of temporal graph data.} \exchange{, including the one forward pass time complexity for various neural network-based models}{More specifically, we compute the time complexity of a forward pass on the entire set of the graph nodes in one snapshot for each method}.

\subsection{GraphERT:}

GraphERT is a Transformer-based model for temporal graphs. It uses \exchange{}{multiple} random walks \exchange{}{with different transition parameters $p$ and $q$} to capture the neighborhood structure around each node at specific time steps. These random walks are fed into a Transformer, which learns node-to-node interactions and their temporal relevance using multi-head attention.

\paragraph{Random Walks Generation:}

For each graph snapshot, the algorithm generates \(\gamma \times n \times |p| \times |q|\) random walks, where:
\begin{itemize}
    \item \(\gamma\) is the number of random walks starting from each node\exchange{}{ for each pair of values assigned to $p$ and $q$}.
    \item \(n\) is the number of nodes in the graph.
    \item \(|p|\) and \(|q|\) are the number of different values for the hyperparameters \(p\) and \(q\).
\end{itemize}

The time complexity for generating the random walks is:

\[
\mathcal{O}(\gamma \times n \times |p| \times |q| \times L)
\]

where \(L\) is the length of each random walk.

\paragraph{Transformer Processing:}

Each random walk is processed by the Transformer. The time complexity of the Transformer is dominated by the self-attention mechanism, which scales quadratically with the sequence length and linearly with the number of attention heads.

For each random walk, the time complexity is:

\exchange{
\[
\mathcal{O}(L^2 \times d \times H \times l)
\]
}{
\[
\mathcal{O}(L^2 \times h_{\text{max}} \times H \times k)
\]
}

where:
\begin{itemize}
    \item \(L\) is the random walk length.
    \item \exchange{\(d\) is the embedding dimension.}{\(h_{\text{max}}\) is the maximum dimensionality of the representation vectors used in different transformer layers. In the original implementation of GraphERT we have \(h_{\text{max}} = d\), but in general it can take any value larger than or equal to $d$.}
    \item \(H\) is the number of attention heads.
    \item \(k\) is the number of layers in the Transformer.
\end{itemize}

\paragraph{Total Time Complexity:}

The total number of random walks is \(\gamma \times n \times |p| \times |q|\). Combining the time complexity for random walk generation and Transformer processing, the total time complexity for processing a single graph snapshot is:


\[
\mathcal{O}\big( n \cdot \gamma \cdot |p|\cdot |q| \cdot (L + L^2 \cdot \exchange{d}{h_{\text{max}}} \cdot H \cdot k) \big) \exchange{}{\in \mathcal{O}\big( n \cdot \gamma \cdot |p|\cdot |q| \cdot (L^2 \cdot h_{\text{max}} \cdot H \cdot k) \big)}
\]

\exchange{}{We can assume that $\gamma$, $|p|$, $q$, $H$ and $k$ are constant values, because they can be fixed values, independent of the graph size ($n$) and the intended dimensionality of the final representations ($d$). Therefore, we can simplify the total complexity as follows:
\[
\mathcal{O}\big( (\gamma \cdot |p|\cdot |q| \cdot H \cdot k) \cdot n \cdot L^2 \cdot h_{\text{max}} \big) \in \mathcal{O}\big( n \cdot L^2 \cdot h_{\text{max}} \big)
\]
}

\exchange{}{
However, it is worth noting that the constant value of this running time is large enough to make practical issues in real experiments. That is why GraphERT shows the most time complexity in Fig. \ref{TV_embeddings}.c. Look at Table \ref{table:GraphERT-HyperParams} for more details about the used values for the hyperparameters of this method.
}

\title{Hyperparameters for GraphERT, DynAE, DynRNN, DynAERNN, and DynGEM}
\author{}
\date{}
\maketitle

\begin{table}[h!]
\centering
\begin{tabular}{l l p{6.8cm}}
\toprule
Method & Hyperparameter & Description / Value \\ 
\midrule
\multirow{8}{*}{GraphERT} 
  & $p$ (Return parameter) & Bias for random walks to return to the previous node, $\in \{0.25, 0.5, 1, 2, 4\}$  \\  
  & $q$ (In-out parameter) & Bias for random walks to explore outward, $\in \{0.25, 0.5, 1, 2, 4\}$ \\  
  & Random Walk Length ($L$) & Length of each random walk (32) \\  
  & Number of Random Walks ($\gamma$) & Number of random walks per node (10) \\  
  & Embedding Dimension ($d$) & Size of node embeddings (8) \\  
  & Attention Heads ($H$) & Number of attention heads (4) \\  
  & Transformer Layers ($k$) & Number of Transformer layers (6) \\  
  & Learning Rate & Learning rate for the Adam optimizer (1e-4) \\  
\bottomrule
\end{tabular}
\caption{Hyperparameters for GraphERT}
\label{table:GraphERT-HyperParams}
\end{table}

\subsection{VGAE:}

To compute the time complexity of a Variational Graph Autoencoder (VGAE) with \(n\) nodes, \(e\) edges, \(k\) Graph Convolutional Network (GCN) layers, and hidden dimensions \(h_1, h_2, \dots, h_k\), where the final latent representation dimension is \(d\), we need to analyze the time complexity at each layer of the GCN. This will account for both node-wise and edge-wise operations.

\paragraph{Step 1: GCN Layer Operations}

A GCN layer applies a linear transformation followed by neighborhood aggregation. The complexity of a single GCN layer is typically determined by:

\begin{itemize}
    \item \textbf{Node-wise operations}: These involve multiplying the node features by a weight matrix. This has a time complexity of \(\mathcal{O}(n \cdot h_{\text{in}} \cdot h_{\text{out}})\), where \(h_{\text{in}}\) is the input dimension of the layer and \(h_{\text{out}}\) is the output dimension.
    \item \textbf{Edge-wise operations}: These involve aggregating the features of neighboring nodes through a message-passing operation over edges. This has a time complexity of \(\mathcal{O}(e \cdot h_{\text{out}})\).
\end{itemize}

\paragraph{Step 2: Time Complexity of Each GCN Layer}

For the \(i\)-th GCN layer:
\begin{itemize}
    \item Let the input feature dimension be \(h_{i-1}\) and the output feature dimension be \(h_i\).
    \item Node-wise multiplication has complexity \(\mathcal{O}(n \cdot h_{i-1} \cdot h_i)\).
    \item Edge-wise aggregation has complexity \(\mathcal{O}(e \cdot h_i)\).
\end{itemize}

Thus, the total time complexity of the \(i\)-th layer is:

\[
\mathcal{O}(n \cdot h_{i-1} \cdot h_i + e \cdot h_i)
\]

\paragraph{Step 3: Summing Over All GCN Layers}

We have \(k\) GCN layers with dimensions \(h_0, h_1, \dots, h_k\), where \(h_0 = n\) is the input feature dimension and \(h_{k} = d\) is the output dimension. Therefore, the total time complexity for all layers is:

\[
T_{\text{GCN}} = \sum_{i=1}^{k} \left( \mathcal{O}(n \cdot h_{i-1} \cdot h_i + e \cdot h_i) \right)
\]

\paragraph{Step 4: VGAE Encoder and Decoder}

\begin{itemize}
    \item \textbf{Encoder}: The encoder, which maps node features to a latent representation space (mean and variance for the latent variables), has the same complexity as the GCN layers, so its complexity is \(T_{\text{GCN}}\).
    \item \textbf{Decoder}: In VGAE, the decoder typically involves reconstructing the adjacency matrix from the latent space. The reconstruction (e.g., using a dot product between latent vectors) has a time complexity of \(\mathcal{O}(n^2 \cdot d)\), as it involves calculating pairwise similarities between all node pairs.
\end{itemize}

\paragraph{Step 5: Total Time Complexity of VGAE}

Summing up the time complexity of the GCN-based encoder and the decoder, we get the overall time complexity:

\[
T_{\text{VGAE}} = T_{\text{GCN}} + \mathcal{O}(n^2 \cdot d)
\]

This expands to:

\[
T_{\text{VGAE}} = \sum_{i=1}^{k} \left( \mathcal{O}(n \cdot h_{i-1} \cdot h_i + e \cdot h_i) \right) + \mathcal{O}(n^2 \cdot d)
\]

\paragraph{Conclusion}

\exchange{So, since the $h_i$ for $i\geq1$ and $d$ are within the same order, The}{Let us denote $\displaystyle \max_{i=1}^k h_i$ by $h_{\text{max}}$. We know that $n = h_0 \ge h_1 \ge \ldots \ge h_k = d$. So, $h_{\text{max}} = h_1$ and the} time complexity of the VGAE is:

\[
T_{\text{VGAE}} = \mathcal{O}\left( \sum_{i=1}^{k} (n \cdot h_{i-1} \cdot h_i + e \cdot h_i) + n^2 \cdot d \right) \in \mathcal{O} \big(n^2\cdot \exchange{d}{h_{\text{max}}}\big) 
\]
\[s.t. ~h_0 = n , h_{k}=d\]

This reflects the complexities of both the encoder (GCN layers) and the decoder (adjacency matrix reconstruction). The most significant term depends on the number of nodes, \exchange{edges,}{} and the dimensions of the latent space. \exchange{}{Hyperparameters of the VGAE model and the values assigned to them in the original paper are listed in Table \ref{table:GraphERT-HyperParams}.}

\begin{table}[h!]
\centering
\begin{tabular}{l l p{6.5cm}}
\toprule
Method & Hyperparameter & Description / Value \\ 
\midrule
\multirow{4}{*}{VGAE}  
  & Latent Dimension ($d$)  & Size of the latent space (dimension of node embeddings) (8) \\ 
  & Graph Convolutional Layers (GCN)  & Number of convolution layers to capture graph structure (2 layers) \\  
  & Learning Rate  & Learning rate for the Adam optimizer  (1e-2) \\  
  & Hidden Dimension ($h$)  & Number of hidden units in the encoder GCN layers (32)  \\  
\bottomrule
\end{tabular}
\caption{Hyperparameters for Variational Graph Autoencoder (VGAE)}
\label{table:VGAE-HyperParams}
\end{table}

\subsection{DynGEM:}

DynGEM uses a \exchange{deep}{Multi-Layer Perceptron (MLP)} autoencoder to generate low-dimensional embeddings for dynamic graphs\exchange{}{ at each snapshot}. At time step \(t=1\), the model is trained on the first snapshot of the graph using a randomly initialized deep autoencoder. For subsequent time steps, embeddings and network parameters are initialized from the previous time step.

Given \(n\) nodes, \(k\) hidden layers with sizes \(h_1, h_2, \dots, h_k\), and the latent representation dimension \(d\), the time complexity of processing \exchange{the first snapshot}{the input graph for each snapshot} is:

\[
\mathcal{O}(n \cdot (n \cdot h_1 + h_1 \cdot h_2 + \dots + h_{k-1} \cdot h_k + h_k \cdot d))
\]

\paragraph{Conclusion}

\exchange{So, since here the $h_i$ for $i\geq1$ and $n$ are within the same order, The}{Let us denote $\displaystyle \max_{i=1}^{k+1} h_i$ by $h_{\text{max}}$. We know that $n = h_0 \ge h_1 \ge \ldots \ge h_{k+1} = d$. So, $h_{\text{max}} = h_1$ and the} time complexity of the DynGEM is:

\[
T_{\text{DynGem}} = \mathcal{O}\left( \sum_{i=1}^{k+1} (n \cdot h_{i-1} \cdot h_i)  \right) \in \mathcal{O} \big(\exchange{n^3}{n^2 \cdot h_{\text{max}}}\big)
\]
\[s.t. ~h_0 = n , h_{k+1}=d\]

\exchange{}{Hyperparameters of this method and the assigned values to them can be found in Table \ref{table:DynGEM-HyperParams}.}

\begin{table}[h!]
\centering
\begin{tabular}{l l p{5.4cm}}
\toprule
Method & Hyperparameter & Description / Value \\ 
\midrule
\multirow{7}{*}{DynGEM}  
  & Latent Dimension ($d$) & Size of the latent space (dimension of node embeddings) (8) \\  
  & Number of layers in the encoder/decoder & Autoencoder consists of 3 layers \\  
  & Layer Sizes ($h_1,h_2$) & Size of each layer in the autoencoder (500, 300) \\  
  & L1 regularization coefficient ($\nu_1$) & Encourages sparsity in model weights ($1e-6$) \\  
  & L2 regularization coefficient ($\nu_2$) & Encourages small weight values ($1e-6$) \\  
  & Learning Rate & Learning rate ($1e-4$) \\  
  & Reconstruction Loss Weight ($\beta$) & Weight for adjacency matrix reconstruction (5) \\  
\bottomrule
\end{tabular}
\caption{Hyperparameters for DynGEM}
\label{table:DynGEM-HyperParams}
\end{table}

\subsection{DynAE:}

DynAE extends a static \exchange{}{MLP} autoencoder to handle temporal graphs. It uses \(l\) look-back adjacency matrices from past snapshots and feeds them into a deep autoencoder to reconstruct the current graph based on previous graphs.

Given an input size of \(n \cdot l\) (where \(n\) is the number of nodes and \(l\) is the number of \exchange{}{leook-back} snapshots), and \(k\) layers in the autoencoder, with the latent representation dimension \(d\), the time complexity for the encoder is:

\[
\mathcal{O}(n \cdot (n \cdot l \cdot h_1 + h_1 \cdot h_2 + \dots + h_k \cdot d)
\]
\paragraph{Conclusion}

\exchange{So, since here the $h_i$ for $i\geq1$ and $n$ are within the same order}{Let us denote $\displaystyle \max_{i=1}^{k+1} h_i$ by $h_{\text{max}}$. We know that $n.l = h_0 \ge h_1 \ge \ldots \ge h_{k+1} = d$. So, $h_{\text{max}} = h_1$. In addition, $l$ can be considered as a constant number}, and the time complexity of the DynAE is:

\[
T_{\text{DynAE}} = \mathcal{O}\left( \sum_{i=1}^{k+1} (n \cdot h_{i-1} \cdot h_i)  \right) \in \mathcal{O} \big(\exchange{n^3}{n^2 \cdot h_{\text{max}}}\big)
\]
\[s.t. ~h_0 = n\cdot l , h_{k+1}=d\]

\exchange{}{Hyperparameters of this method and the assigned values to them can be found in Table \ref{table:DynAE-HyperParams}.}

\begin{table}[h!]
\centering
\begin{tabular}{l l p{5.7cm}}
\toprule
Method & Hyperparameter & Description / Value \\ 
\midrule
\multirow{8}{*}{DynAE}  
  & Look-back ($l$) & Number of previous snapshots used (2) \\  
  & Latent Dimension ($d$) & Size of the latent space (dimension of node embeddings) (8) \\  
  & Number of layers in the encoder/decoder & Autoencoder consists of 3 layers \\  
  & Layer Sizes ($h_1,h_2$) & Size of each autoencoder layer (500, 300) \\  
  & L1 regularization coefficient ($\nu_1$) & Encourages sparsity in model weights ($1e-6$) \\  
  & L2 regularization coefficient ($\nu_2$) & Encourages small weight values ($1e-6$) \\  
  & Learning Rate & Learning rate ($1e-4$) \\  
  & Reconstruction Loss Weight ($\beta$) & Weight for adjacency matrix reconstruction (5) \\  
\bottomrule
\end{tabular}
\caption{Hyperparameters for DynAE}
\label{table:DynAE-HyperParams}
\end{table}

\subsection{DynRNN:}

DynRNN \exchange{}{is similar to DynAE, but it} uses Recurrent Neural Networks (RNNs), specifically Long Short-Term Memory (LSTM) networks, to capture temporal dependencies across snapshots. Each node’s neighborhood at each snapshot is passed into the LSTM.

\exchange{Given \(n\) nodes, \(k_{LSTM}\) LSTM layers with sizes \(h_{1_{LSTM}}, h_{2_{LSTM}}, \dots, {h_{k_{LSTM}}}\) and \(l\) look-back snapshots, t}{T}he time complexity for LSTM step $i$ \exchange{}{on one node} is:

\[
\mathcal{O}(h_{i-1_{LSTM}} \cdot h_{i_{LSTM}}+ h_{i_{LSTM}}^2)
\]

\exchange{Now, for the entire graph,}{Given \(n\) nodes, \(k_{LSTM}\) LSTM layers with sizes \(h_{1_{LSTM}}, h_{2_{LSTM}}, \dots, {h_{k_{LSTM}}}\) and \(l\)  snapshots,}
 the total time complexity for one snapshot is:


\[
\mathcal{O}\big( n\cdot(n\cdot l \cdot h_{1_{LSTM}} + h_{1_{LSTM}} \cdot h_{2_{LSTM}}+ \cdots + h_{k-1_{LSTM}} \cdot h_{k_{LSTM}}+ h_{k_{LSTM}}\cdot d +  h_{1_{LSTM}}^2+\cdots + h_{k_{LSTM}}^2 + d^2 ) \big)
\]
\paragraph{Conclusion}

\exchange{So, since here the $h_{i_{LSTM}}$ for $i \geq 1$ and $n$ are within the same order}{Let us denote $\displaystyle \max_{i=1}^{k+1} h_{i_{LSTM}}$ by $h_{\text{max}}$. We know that $n \cdot l = h_{0_{LSTM}} \ge h_{1_{LSTM}} \ge \ldots \ge h_{k+1_{LSTM}} = d$. So, $h_{\text{max}} = h_{1_{LSTM}}$. Is addition, $l$ can be considered as a constant number}, the time complexity of the DynRNN is:

\[
T_{\text{DynRNN}} = \mathcal{O}\left( \sum_{i=1}^{k+1} (n \cdot (h_{i-1_{LSTM}} \cdot h_{i_{LSTM}}+h^2_{i_{LSTM}}))  \right) \in \mathcal{O} \big(\exchange{n^3}{n^2 \cdot h_{\text{max}}}\big)
\]
\[s.t. ~h_{0_{LSTM}} = n\cdot l , h_{k+1_{LSTM}}=d\]

\exchange{}{Hyperparameters of this method and the assigned values to them can be found in Table \ref{table:DynRNN-HyperParams}.}

\begin{table}[h!]
\centering
\begin{tabular}{l l p{6.5cm}}
\toprule
Method & Hyperparameter & Description / Value \\ 
\midrule
\multirow{8}{*}{DynRNN}  
  & Look-back ($l$) & Number of previous snapshots used (2) \\  
  & Latent Dimension ($d$) & Size of the latent space (dimension of node embeddings) (8) \\  
  & Number of RNN Layers & Number of stacked LSTM layers (3) \\  
  & Hidden State Size & Number of hidden units in LSTM (500, 300) \\  
  & L1 regularization coefficient ($\nu_1$) & Encourages sparsity in model weights ($1e-6$) \\  
  & L2 regularization coefficient ($\nu_2$) & Encourages small weight values ($1e-6$) \\  
  & Learning Rate & Learning rate ($1e-4$) \\  
  & Reconstruction Loss Weight ($\beta$) & Weight for adjacency matrix reconstruction (5) \\  
\bottomrule
\end{tabular}
\caption{Hyperparameters for DynRNN}
\label{table:DynRNN-HyperParams}
\end{table}

\subsection{DynAERNN:}

DynAERNN combines the autoencoder from DynAE with the LSTM-based RNN from DynRNN. The encoder compresses the neighborhood vectors \exchange{}{of $l$ snapshots} into a low-dimensional space, which the LSTM processes across time to capture temporal dependencies.

The total time complexity for DynAERNN is the sum of the autoencoder and LSTM complexities:

\begin{multline*}
\mathcal{O}(n \cdot (n \cdot l \cdot h_1 + h_1 \cdot h_2 + \dots + h_{k-1} \cdot h_{k}) + \\\mathcal{O}\big( n\cdot(h_k \cdot h_{1_{LSTM}} + h_{1_{LSTM}} \cdot h_{2_{LSTM}}+ \cdots + h_{k-1_{LSTM}} \cdot h_{k_{LSTM}}+ h_{k_{LSTM}}\cdot d +  h_{1_{LSTM}}^2+\cdots + h_{k_{LSTM}}^2 + d^2 ) \big)
\end{multline*}
\paragraph{Conclusion}

\exchange{So, since here the $h_i$ and $h_{i_{LSTM}}$ for $i \geq 1$ as well as $n$ are within the same order, The}{Let us denote $\displaystyle \max(\max_{i=1}^k h_i , \max_{i=1}^{k+1} h_{i_{LSTM}})$ by $h_{\text{max}}$. We know that $n \cdot l = h_0 \ge h_1 \ge \ldots \ge h_k = h_{0_{LSTM}} \ge h_{1_{LSTM}} \ge \ldots \ge h_{k+1_{LSTM}} = d$. So, $h_{\text{max}} = h_1$. In addition, $l$ can be considered as a constant number} time complexity of the DynRNN is:

\[
T_{\text{DynAERNN}} = \mathcal{O}\left( \sum_{i=1}^{k} (n \cdot h_{i-1} \cdot h_i) + \sum_{i=1}^{k+1} (n \cdot (h_{i-1_{LSTM}} \cdot h_{i_{LSTM}}+h^2_{i_{LSTM}}))  \right) \in \mathcal{O} \big(\exchange{\n^3}{n^2 \cdot h}\big)
\]
\[s.t. ~h_0 = n\cdot l , h_{0_{LSTM}}=h_k, h_{k+1_{LSTM}}=d\]

\exchange{}{Hyperparameters of this method and the assigned values to them can be found in Table \ref{table:DynAERNN-HyperParams}.}

\begin{table}[h!]
\centering
\begin{tabular}{l l p{6cm}}
\toprule
Method & Hyperparameter & Description / Value \\ 
\midrule
\multirow{9}{*}{DynAERNN}  
  & Look-back ($l$) & Number of previous snapshots used (2) \\  
  & Latent Dimension ($d$) & Size of the latent space (dimension of node embeddings) (8) \\  
  & Autoencoder Layer Sizes & Size of each autoencoder layer (500, 300) \\  
  & Number of RNN Layers & Number of stacked LSTM layers (3) \\  
  & LSTM Hidden State Size & Number of hidden units in LSTM (500, 300) \\  
  & L1 regularization coefficient ($\nu_1$) & Encourages sparsity in model weights ($1e-6$) \\  
  & L2 regularization coefficient ($\nu_2$) & Encourages small weight values ($1e-6$) \\  
  & Learning Rate & Learning rate ($1e-4$) \\  
  & Reconstruction Loss Weight ($\beta$) & Weight for adjacency matrix reconstruction (5) \\  
\bottomrule
\end{tabular}
\caption{Hyperparameters for DynAERNN}
\label{table:DynAERNN-HyperParams}
\end{table}

\subsection{\TV{}:}

The time complexity of the \TV{} framework is driven by several components, including GNN layers, GRU operations, and an attention mechanism. Below, we break down \exchange{}{the total complexity into} the time complexity \exchange{for}{of} each component.

\paragraph{1. GNN and GRU Layers:}

At each time step \(t\), the model processes the graph using a combination of GNN layers and a GRU-based RNN. The time complexity for these operations can be broken down as follows:

\begin{itemize}
    \item \exchange{}{ \textbf{Low-dimensional Embedding}: first of all, each $n$-dimensional neighborhood vector is mapped to a $h_{GRU}$-dimensional embedding using a one layer feed forward network. The time complexity of this part will be:
    \[
    \mathcal{O}(n^2 \cdot h_{GPU})
    \]
    }
    \item \textbf{Graph Convolution (GNN)}: Similar to the VGAE mentioned above , the time complexity of the GNN layer is:
    \[
T_{\text{GNN}} = \sum_{i=1}^{k} \left( \mathcal{O}(n \cdot h_{i-1} \cdot h_i + e \cdot h_i) \right)
\]
    \item \textbf{GRU Operation}: \exchange{Each GRU cell performs operations with two main gates: the reset gate and the update gate. The time complexity for GRU operations per time step is:}{Since the inner functions of our GPU cell is implemented by GCN layers, the dominant term in the time complexity of the GPU cell in each time step is equal to:}
    \[
    \mathcal{O}(\exchange{n \cdot (h_{GRU}^2 + h_{GRU})}{n \cdot h_{GRU}^2 + e \cdot h_{GRU}})
    \]
\end{itemize}

\paragraph{2. Temporal Attention Mechanism:}

The attention mechanism aggregates past hidden states over a window of size \(w\). \exchange{The time complexity for computing attention weights is:}{The attention of the model into the last $w$ snapshots is computed in:} 

\[
\mathcal{O}(w \cdot h)
\]

where \(w\) is the attention window size and \(h\) is the hidden dimension. \exchange{}{The time complexity of computing the weighted average vectors for all the $n$ node according to these computed attentions is:
\[
\mathcal{O}(n\cdot w \cdot h)
\]}

\paragraph{\exchange{}{3. Reconstruction:}}
\exchange{}{Similar to VGAE, the reconstruction process in \TV{} is through computing the inner product of the final representation of each pair of the nodes, and its time complexity is:
\[
\mathcal{O}(n^2 \cdot d)
\]}

\paragraph{4. Overall Time Complexity for Each Time Step:}

The overall time complexity at each time step is a combination of the \exchange{}{initial projection to a low-dimensional space using a feedforward layer,} GNN and GRU computations, attention mechanism, \exchange{reconstruction and feedforward layers:}{and reconstruction:}

\[
\mathcal{O}(n \cdot (h_1 + h_1 \cdot h_2 + \dots + h_k \cdot d) + e \cdot (h_1 + \cdots + h_k) + \exchange{n \cdot (h_{GRU}^2 + h_{GRU})}{n \cdot {h_{GRU}^2} + e \cdot h_{GRU}} + (n+1) \cdot w \cdot h + n^2 \cdot d)
\]

\paragraph{Conclusion}

\exchange{Since the $h_i$ for $i\geq1$, $h_{GRU}$,and $d$ are within the same order, The}{{Let us denote $\displaystyle \max(\max_{i=1}^{k+1} h_i , h_{\text{GRU}}, h)$ by $h_{\text{max}}$. We know that $n \cdot l = h_0 \ge h_1 \ge \ldots \ge h_k+1 = d$. So, $h_{\text{max}} = h_1$. We can infer that the}} time complexity of \exchange{the}{} \TV{} is:

\begin{eqnarray}
T_{\text{\TV{}}} = &\mathcal{O}\left( \sum_{i=1}^{k+1} (n \cdot h_{i-1} \cdot h_i + e \cdot h_i) + \exchange{n \cdot (h_{GRU}^2 + h_{GRU})}{n \cdot h_{GRU}^2 + e \cdot h_{GRU}} + n \cdot w \cdot h + n^2 \cdot d\right) \in \mathcal{O} \big(n^2\cdot \exchange{d}{h_{\text{max}} + n \cdot w \cdot h}\big) 
\nonumber \\
&s.t. ~h_0 = 1 , h_{k+1}=d \nonumber
\end{eqnarray}

The summary of the time complexities \exchange{}{for different methods} is shown in Table \ref{full:time_complexity}.

\begin{table}
\centering
\caption{One forward pass time complexity for one time window (i.e. snapshot).}
\label{full:time_complexity}
\begin{tabular}{c|c}
\hline
\textbf{Method} & \textbf{Complexity} \\
\hline
VGAE & $\mathcal{O}\left( \sum_{i=1}^{k} (n \cdot h_{i-1} \cdot h_i + e \cdot h_i) + n^2 \cdot d \right) \in \mathcal{O} \big(n^2\cdot \exchange{d}{h_{\text{max}}}\big)$ \\
\hline
DynGEM & $\mathcal{O}\left( \sum_{i=1}^{k+1} (n \cdot h_{i-1} \cdot h_i) \right) \in \mathcal{O} \big(\exchange{n^3}{n^2 \cdot h_{\text{max}}}\big)$ \\
\hline
DynAE & $\mathcal{O}\left( \sum_{i=1}^{k+1} (n \cdot h_{i-1} \cdot h_i) \right) \in \mathcal{O} \big(\exchange{n^3}{n^2 \cdot h_{\text{max}}}\big)$ \\
\hline
DynRNN & $\mathcal{O}\left( \sum_{i=1}^{k+1} (n \cdot (h_{i-1_{\text{LSTM}}} \cdot \exchange{h_i}{h_{i_{\text{LSTM}}}} + h^2_{i_{\text{LSTM}}})) \right) \in \mathcal{O} \big(\exchange{n^3}{n^2 \cdot h_{\text{max}}}\big)$ \\
\hline
DynAERNN & $\mathcal{O}\left( \sum_{i=1}^{k} (n \cdot h_{i-1} \cdot h_i) + \sum_{i=1}^{k+1} (n \cdot (h_{i-1_{\text{LSTM}}} \cdot \exchange{h_i}{h_{i_{\text{LSTM}}}} + h^2_{i_{\text{LSTM}}})) \right) \in \mathcal{O} \big(\exchange{n^3}{n^2 \cdot h_{\text{max}}}\big)$ \\
\hline
GraphERT & \exchange{$\mathcal{O}\big( n \cdot \gamma \cdot |p| \cdot |q| \cdot (L + L^2 \cdot d \cdot H \cdot k) \big)$}{$\mathcal{O}\big( (\gamma \cdot |p|\cdot |q| \cdot H \cdot k) \cdot n \cdot L^2 \cdot h_{\text{max}} \big) \in \mathcal{O}\big( n \cdot L^2 \cdot h_{\text{max}} \big)
$} \\
\hline
\TV{} & $\mathcal{O}\left( \sum_{i=1}^{k+1} (n \cdot h_{i-1} \cdot h_i + e \cdot h_i) + \exchange{n \cdot (h_{\text{GRU}}^2 + h_{\text{GRU}})}{n \cdot h_{\text{GRU}}^2 + e \cdot h_{\text{GRU}}} + \exchange{}{n \cdot} w \cdot h + n^2 \cdot d  \right) \in \mathcal{O} \big(n^2 \cdot \exchange{d}{h_{\text{max}} + n \cdot w \cdot h_{\text{max}}}\big)$ \\
\hline
\end{tabular}
\end{table}

\section{Additional Results}
\label{add_results}
We present the learned representations of the three best
performing windows in terms of the culture's hit/miss ratios during Gameplay for two additional cultures in Extended Data Figs. \ref{game_embeddings_9353} and \ref{game_embeddings_11614}. The figures repeatedly demonstrate \TV{}'s outperformance over the other baseline methods in identifying clusters of channels that belong to the same region on the HD-MEA.


\end{document}